\documentclass[aps,prd,twocolumn,superscriptaddress,groupedaddress,nofootinbib]{revtex4-1}  

\usepackage{graphicx}   % needed for figures
\usepackage{dcolumn}   % needed for some tables
\usepackage{bm}            % for math
\usepackage{amssymb}  % for math
\usepackage{amsmath}
\usepackage{blindtext}
\usepackage{comment}

\numberwithin{equation}{section}  % section numbering of equations

\newcommand\blfootnote[1]{%
  \begingroup
  \renewcommand\thefootnote{}\footnote{#1}%
  \addtocounter{footnote}{-1}%
  \endgroup
}

\begin{document}

\hphantom\\
\begin{flushright}
${}$\\[-2cm]
MAN/HEP/2018/03\\
August 2018
\end{flushright}
\vspace{-7mm}

\title{\Large{Protecting the Stability of the EW~Vacuum}\\
	\Large{from Planck-Scale Gravitational Effects}}

%\date{\today}

\author{ Vincenzo~Branchina$^{\,a,b\,*}$, Filippo~Contino$^{\,a,c\,\dagger}$
and Apostolos~Pilaftsis$^{\,d\,\ddagger}$\vspace{2mm}}

\affiliation{\small $^a$Department of Physics and Astronomy, University of 
Catania, \\ Via Santa Sofia 64, 95123 Catania, Italy\vspace{1mm}}

\affiliation{\small $^b$INFN, Sezione di Catania, Via Santa Sofia 64,
  95123 Catania, Italy\vspace{1mm}} 

\affiliation{\small $^c$Scuola Superiore  di Catania, Via Valdisavoia
  9, 95123 Catania, Italy\vspace{1mm}} 

\affiliation{\small $^d$Consortium for Fundamental Physics, School of Physics and Astronomy,\\ 
University of Manchester, Manchester, M13 9PL, United Kingdom\vspace{1mm}}

\begin{abstract}
\centerline{\bf ABSTRACT}
\medskip
\noindent
We investigate the stability of the Standard-Model Electroweak (EW)
vacuum in the presence of Planck-scale suppressed operators of the
type~$\phi^{2n}/M^{2n-4}_{\rm P}$ that involve the Higgs field~$\phi$
and could in principle be induced by quantum gravity effects. We
show how minimal embeddings of the Standard Model~(SM) in supergravity
(SUGRA) can stabilize the EW vacuum against such operators up to very
high values of the induced\- super\-symmetry breaking scale~$M_{\cal S}$,
which may well be above the onset of the so-called SM meta\-stability
scale of~$10^{11}$~GeV.  In particular, we explicitly demonstrate how
discrete\- $R$ symmetries could be invoked to suppress the occurrence
of harmful Planck-scale operators\- of the
form~$\phi^{2n}/M^{2n-4}_{\rm P}$ to arbitrary higher powers of~$n$.
We~analyze different scenarios of Planck-scale gravitational physics
and derive\- lower limits on the power $n$ that is required in order to
protect our EW vacuum from dangerous\- rapid decay. The significance
of our results for theories of low-scale quantum gravity is
illustrated.
\end{abstract}

\maketitle

\blfootnote{$^*$E-mail: {\tt Branchina@ct.infn.it}\\
$^\dagger$E-mail: {\tt filippo.contino@ct.infn.it}\\
$^\ddagger$E-mail:  {\tt Apostolos.Pilaftsis@manchester.ac.uk} }
\vspace{-1.1cm}

\section{Introduction}\label{sec:intro}
\setcounter{equation}{0}

The problem of stability of the electroweak (EW)
vacuum~\cite{Cabibbo:1979ay,Flores:1982rv,Lindner:1986,Sher:1988mj,Lindner:1988ww,Ford:1992mv,Sher:1993mf,Casas:1996aq,Casas:1994qy,Altarelli:1994rb,Isidori:2001bm,Espinosa:2007qp,Lee:2008hz,Lee:2012ug,Lee:2014ula}
has been central not only in our understanding of the Standard
Model~(SM), but also in demystifying the very nature of possible New
Physics~(NP). In particular, recent measurements of the Higgs-boson and the
top-quark masses seem to indicate that we live in a metastable
Universe, in which the true, absolutely stable vacuum is not the one
we now live in, of order 100 GeV, but at scales much higher than this,
typically~larger than $10^{11}$ GeV. {\em Why do we then still exist?}

Earlier studies that attempted to address this simple but fundamental
question were primarily focused on establishing bounds for the Higgs
boson mass~$M_H$. Thus far, three different directions have been
followed in the literature: (i)~the EW vacuum $v \approx 245$~GeV was
considered to be absolutely stable, by requiring that the Higgs
effective potential $V(\phi)$ does not take values lower than that of
the EW minimum; (ii)~the EW vacuum has become a metastable state
representing a relative minimum of $V(\phi)$, but its lifetime turns
out to be larger than the age of the
Universe~\cite{Turner:1982xj,Hut:1983xa,Flores:1982rv,Isidori:2001bm};
(iii)~any possible higher-scale minimum of $V(\phi)$ happens to be
degenerate with that of the EW vacuum, obeying some {\it ad~hoc}
principle of {\em multiple
	criticality}~\cite{Froggatt:1995rt}. Postulating the latter
principle enabled the authors of~\cite{Froggatt:1995rt} to obtain
predictions for the masses of the Higgs boson $H$ and the top quark
$t$, well before their discovery. The so-predicted masses came out to
be surprisingly close to their current experimental {\em central}
values:~$M_t \simeq 173.34$~GeV and $M_H \simeq
125.09$~GeV~\cite{Aad:2015zhl,ATLAS:2014wva}, within the expected theoretical
errors~\cite{Bednyakov:2015sca}.

In the context of Quantum Field Theory (QFT), the decay rate of a
metastable state (false vacuum) was first calculated by Coleman and
Callan, who studied the decay in a flat spacetime
background~\cite{Coleman:1977py,Callan:1977pt}, and subsequently by
Coleman and De Luccia, who included the effect of
gravity~\cite{Coleman:1980aw}.  The decay is triggered by quantum
fluctuations that induce a finite probability for a bubble created at
the true vacuum $\phi_{\rm tv}$ to materialize in a false vacuum
($\phi_{\rm fv}$) sea. Coleman and collaborators considered a scalar
theory where the potential $V(\phi)$ has a relative and an absolute
minimum at $\phi_{\rm fv}$ and $\phi_{\rm tv}$, respectively, with an
energy density difference $V(\phi_{\rm fv}) - V(\phi_{\rm tv})$ much
smaller than the height of the ``potential barrier'',
$V(\phi_{\rm top}) - V(\phi_{\rm fv})$, where $V(\phi_{\rm top})$ is
the maximum of the potential between the two minima.  Given this
condition, the true vacuum bubble is separated from the false vacuum
sea by a ``thin wall'', and this allows to treat the problem
analytically, within the so-called {\em thin-wall approximation}.

With the discovery of the Higgs boson in 2012, the above question
regarding the stability${}$ of the EW vacuum received renewed
interest. The goal now is no longer to derive bounds on $M_H$ as a
function of $M_t$, but rather perform precision analyses beyond the
leading- and next-to-leading-order approximation in an attempt to:
(i)~accurately delineate the interfacial region between absolute
stability and metastability~\cite{Degrassi:2012ry,Garbrecht:2015oea,Grinstein:2015jda,
	Garbrecht:2017idb}, (ii)~study the consequences of the EW-vacuum stability
on the evolution of the early
Universe~\cite{Herranen:2014cua,Khan:2014kba,
	Herranen:2015ima,Kearney:2015vba,
	Anchordoqui:2015fra,Kahlhoefer:2015jma,Ema:2016kpf,Ema:2016ehh,Okada:2015lia,
	Urbanowski:2016pks,Stachowski:2016zpq},
and (iii)~test the impact of the latter on the 
viability of different NP
scenarios~\cite{Mihaila:2012fm,Chetyrkin:2012rz,Bezrukov:2012sa,EliasMiro:2011aa,
	Degrassi:2012ry,Buttazzo:2013uya,Branchina:2013jra,Branchina:2014usa,Branchina:2014rva,
	Branchina:2015nda,Haba:2014oxa,Ferreira:2015pfi,Chakrabarty:2016smc}.

Next-to-next-to-leading order (NNLO) computations of the SM effective
potential~$V(\phi)$ show that $t$-quark loops make $V(\phi)$ to turn
over for sufficiently large field values of $\phi \gg v $. Taking the
central values for $M_t$ and $M_H$ quoted above at face value, one
obtains that a second lower minimum, $\phi_{\rm tv}$, for the
potential $V(\phi)$ gets developed, which is much deeper than the EW
one $v\equiv \phi_{\rm fv} $, for
$\phi_{\rm tv}\, \gg \, \phi_{\rm fv}$.  Here, another relevant scale
is the so-called {\em metastability} or {\em instability}
scale~$\phi_{\rm inst}$~\cite{Cabibbo:1979ay}.  The scale
$\phi_{\rm inst}$ is defined as the value, for which
$V(\phi_{\rm inst}) = V(v) = 0$, where the normalization of $V (v)$ to
a vanishingly small cosmological constant was considered. Hence, for
$\phi > \phi_{\rm inst}$, the potential $V(\phi)$ becomes negative.
For the central values of the Higgs and top masses reported above, one
finds that $\phi_{\rm inst} \sim 10^{11}$~GeV. Moreover, in the SM the
thin-wall approximation,
i.e.~$V(\phi_{\rm fv}) - V(\phi_{\rm tv})\ll V(\phi_{\rm top}) -
V(\phi_{\rm fv})$,
may not hold true in general.  Consequently, the approximate
results of \cite{Coleman:1977py,Callan:1977pt} and
\cite{Coleman:1980aw} may no longer be applicable. In this case, one
is compelled to study this problem numerically.

The decay of the EW vacuum was first studied in a flat spacetime
background, by considering the interesting possibility that the SM is
valid all the way up to the Planck scale
$M_{\rm P} =1.9\times 10^{19}$~GeV, and any effects of NP and quantum
gravity will show up only at this scale. A crucial assumption of this
analysis was that for the calculation of the EW vacuum lifetime
$\tau$, the presence of NP operators suppressed by~$M_{\rm P}$ could
be neglected~\cite{Isidori:2001bm}, and as such, $\tau$ was calculated
by considering SM interactions
only~\cite{Isidori:2001bm,Espinosa:2007qp,Mihaila:2012fm,Chetyrkin:2012rz,Bezrukov:2012sa,EliasMiro:2011aa,Degrassi:2012ry,Buttazzo:2013uya}. The
main reason for this simplification given in~\cite{Espinosa:2007qp}
was that even if Planck-scale suppressed operators of NP were present,
they can still be neglected because the instability scale
$\phi_{\rm inst} \sim 10^{11}$ GeV happens to be orders of magnitude
smaller than $M_{\rm P}$, thereby tacitly assuming the existence of
some kind of a decoupling mechanism.

Restricting ourselves to SM interactions only, the lifetime $\tau$ of
the EW vacuum was found to be much larger than the age of our Universe
$T_U \approx 13.7 \times 10^9$ years~\cite{Degrassi:2012ry,Buttazzo:2013uya},
i.e.~$\tau \sim 10^{640} \, T_U$ (see \cite{Bentivegna:2017qry}). This result is obtained when the
effect of standard minimal gravity on the spacetime background 
metric is ignored. Taking the latter into account, $\tau$ gets
``slightly" modified~\cite{Isidori:2007vm,Branchina:2016bws,Rajantie:2016hkj}:
$\tau \sim 10^{660} \, T_U$ (see \cite{Bentivegna:2017qry}). Moreover, if a non-minimal interaction
$\xi \phi^2 R$ between the Higgs field $\phi$ and the Ricci scalar $R$
is considered, the lifetime $\tau$ will crucially depend on the value of the non-minimal coupling $\xi$. In particular, in the conformal
limit~$\xi \to 1/6$, the result of $\tau$ obtained for a flat
spacetime background is recovered~\cite{Rajantie:2016hkj}.

In most SM predictions for the lifetime~$\tau$ of the EW vacuum, the
working hypothesis was that NP operators appearing at scales close to
$M_{\rm P}$ ($\gg \phi_{\rm inst}$) will decouple and so have no
relevant effect on~$\tau$. However, this hypothesis turns out to be
misleading. It was shown in~\cite{Branchina:2013jra,Branchina:2014usa,Branchina:2014rva,Branchina:2015nda} that the
presence of such NP operators can drastically modify${}$~$\tau$, when
considering a flat spacetime background. On the other hand, it is
expected on general grounds~\cite{Coleman:1980aw} that the inclusion of gravity
will partially counteract the destabilizing effect of any Planckian NP
on the EW vacuum. This prompted the authors of~\cite{Espinosa:2015zoa} to
claim that its opposing effect will be so strong,
so that the original SM predictions for $\tau$
(i.e.~$\tau \sim 10^{660}\,T_U$) will be re-obtained, thus enabling them to
ignore the effect of Planckian NP on~$\tau$ altogether.  Nevertheless${}$, a recent
dedicated analysis does {\em not} corroborate such assertions, and the instability of
the EW vacuum persists even if gravity is taken into account~\cite{Bentivegna:2017qry}.

It has now become more evident than ever that Planckian NP can
strongly affect the stability of the EW vacuum and the presence of
possible harmful Planck-scale-suppressed operators of the form
$\phi^{2n}/M^{2n-4}_{\rm P}$ can no longer be ignored in the
computation${}$. Although we may not be able to exclude {\it a priori}
such harmful operators, one may still wonder whether a protective
symmetry can be invented in order to postpone their appearance to
arbitrarily high orders $n$, so as to render their destabilising
effect on the EW vacuum harmless. Our quest is motivated by
analogous\- protective symmetries that were invoked in axion physics
to suppress Planck-scale effects and so render the Peccei-Quinn
mechanism for solving the strong CP problem
effective~\cite{Kamionkowski:1992mf,Dias:2014osa,DiLuzio:2017tjx}.

In this paper we will show how supergravity (SUGRA) embeddings of the
SM~\cite{Nilles:1983ge} could be sufficient to protect the stability
of the EW vacuum up to very large values of the soft supersymmetry
(SUSY) breaking scale~$M_{\cal S}$, above the so-called SM
metastability scale of~$10^{11}$~GeV. Moreover, we will explicitly
demonstrate how discrete $R$ symmetries could be used in order to
restrict the form of the holomorphic super\-potential~${\cal W}$, and so
suppress the appearance of the harmful Planck-scale operators of the
type $\phi^{2n}/M^{2n-4}_{\rm P}$ to arbitrary higher powers of~$n$.

The layout of the paper is as follows. After this introductory
section, Section~\ref{sec:thback} provides the necessary theoretical
background for our computations of the lifetime~$\tau$ of the EW
vacuum. In the same section, we use this theoretical background to
reaffirm the known predictions for~$\tau$ in the SM in the absence of
any Planckian NP. In Section~\ref{sec:PlanckNP}, we analyze the impact
of the harmful operators of the type $\phi^{2n}/M^{2n-4}$ on~$\tau$,
for Planck-scale NP scenarios with $M = M_{\rm P}$ and
$M = M_{\rm P}/10$.  Section~\ref{sec:SUGRA} discusses minimal
embeddings of the SM in SUGRA, which is a well-motivated theoretical
framework that provides good control of Planck-scale physics.
We~will show how discrete $R$ symmetries could be employed to postpone
the appearance of harmful Planck-scale suppressed operators to higher
powers of~$n$, thus rendering them technically harmless and safe for
the stability of the EW vacuum. In Section~\ref{sec:sspotential}, we
present a numerical analysis of a few representative scenarios with a
low and high soft SUSY-breaking scale $M_{\cal S}$, i.e.~for
$M_{\cal S} = 10$~TeV and $M_{\cal S} = 10^9$~TeV. In all our
numerical estimates, we consider the effect of gravity on the
tunnelling rate from the false EW vacuum to the true and absolutely
stable trans-Planckian vacuum.  Finally, Section~\ref{sec:Concl}
contains our conclusions.

\setcounter{equation}{0}
\section{Theoretical Background} 
\label{sec:thback}

In this section, we will briefly review the theoretical framework
needed for computing the tunnelling time~$\tau$ from a Minkowski
false vacuum $\phi_{\rm fv}$ to an Anti--de Sitter~(AdS) true vacuum
$\phi_{\rm tv}$ in the classical, leading-order
approximation~\cite{Coleman:1977py}. With the aid of this framework,
we will then be able to re-establish the known results for~$\tau$ in
the SM, in the absence of any Planckian NP.

\subsection{General Framework}

We will now describe the general computational framework for
evaluating the lifetime of the EW vacuum~$\tau$ by considering first
an Euclidean flat spacetime, before turning our attention to an
$\mathbb{O}(4)$-symmetric curved background metric.

\subsubsection{Euclidean Flat Spacetime}

Let us first consider the Euclidean flat-spacetime action for a 
real scalar field $\phi$,
\begin{equation}
\label{maction}
S[\phi]\ =\ \int\! d^4x\: \bigg( \frac 1 2 \partial_\mu\phi\, \partial_\mu\phi\: 
+\: V(\phi) \bigg)\;,
\end{equation}
where $V(\phi)$ denotes the potential  with a local minimum 
(\emph{false vacuum}) at $\phi=\phi_{\rm fv}$, and 
an absolute minimum (\emph{true vacuum}) at $\phi=\phi_{\rm tv}$.
Taking into account the observed smallness of the cosmological
constant, we require that $V(\phi_{\rm fv}) = 0$, which is an excellent
approximation for our purposes.

In order to calculate the lifetime of the false vacuum, we have to
first find the so-called \emph{bounce solution}~$\phi_b (r)$ to the
Euler--Lagrange equation of motion for~$\phi$ derived from the
Euclidean action~$S[\phi]$ in~\eqref{maction}. We note that the bounce
solution~$\phi_b (r)$ is $\mathbb{O}(4)$-symmetric and so depends only on
the radial coordinate~$r$. Moreover, it has to satisfy certain
boundary conditions~\cite{Coleman:1977py,Callan:1977pt}. More explicitly, $\phi_b (r)$ must
be a solution to the Euclidean Euler-Lagrange equation given by
\begin{equation}
\label{meq}
\ddot \phi(r)\: +\: \frac{3}{r} \dot \phi (r)\ =\ \frac{d V}{d \phi}\;,
\end{equation}
subject to the boundary conditions,
\begin{equation}
\label{mbc}
\phi(\infty)\ =\ \phi_{\rm fv}\;, \qquad \dot \phi(0)\ =\ 0\;.
\end{equation}
Here and in the following, an overdot indicates a derivative with respect to $r$.

Knowing~$\phi_b (r)$ enables us to evaluate the action~$S$ at $\phi_b$
as follows:
\begin{equation}
\label{action}
S[\phi_b]\ =\ 2 \pi^2 \int_0^\infty \! d r\, r^3\:  
\bigg(\, \frac 1 2 \dot \phi_b^2\: +\: V(\phi_b) \bigg)\; .
\end{equation}
In fact, following a reasoning similar to the one used to prove
Derrick's theorem,  the kinetic term $\frac{1}2\,\dot \phi_b^2$
in~\eqref{action} may effectively be replaced with $-2\,V(\phi_b)$, and
so a simpler form for the action may be derived, i.e.
\begin{equation}
\label{eq:Sflat}
S[\phi_b]\ =\ -\,2 \pi^2 \int_0^\infty \! d r\, r^3 \, V(\phi_b)\; .
\end{equation}

We now have all the ingredients to evaluate the decay rate $\Gamma$ of
the false vacuum, which is given by
\begin{equation} 
\label{gamma}
\Gamma\ =\ D\, \exp\!\Big[\! -\big(S[\phi_b]-S[\phi_{\rm fv}]\big)\Big]\
\equiv\ D\, \exp\!\big(\!-\!B\big)\;,
\end{equation}
where $B\equiv S[\phi_b]-S[\phi_{\rm fv}]$ is usually termed the {\em
	tunnelling exponent}. The negative of the tunnelling exponent, $-B$,
determines the leading-order contribution to the decay
rate~$\Gamma \equiv 1/\tau$, while $D$ is known as the {\em quantum
	fluctuation determinant} to be discussed below~$^{1}$ \footnotetext[1]{The recent
	renewed interest for the vacuum stability problem prompted a more
	careful treatment of issues, such as the gauge invariance of the vacuum
	decay rate and the contribution of zero modes to the quantum
	fluctuation determinant
	\cite{Andreassen:2014gha,DiLuzio:2014bua,Andreassen:2014eha,Endo:2017tsz,Andreassen:2017rzq,Chigusa:2017dux}.}.
Given that
$V(\phi_{\rm fv})=0$, the action $S[\phi_{\rm fv}]$ vanishes, and the
tunnelling exponent becomes $B=S[\phi_b]$.  In order to determine the
lifetime~$\tau \equiv 1/\Gamma$ of the false vacuum in the flat
spacetime, we first solve numerically~\eqref{meq} with boundary
conditions~\eqref{mbc}, and then use~\eqref{eq:Sflat} to get the tunnelling
exponent~$B$ of~\eqref{gamma}.

\subsubsection{{\boldmath $\mathbb{O}(4)$}-Symmetric Curved Spacetime}

The next step is to study the impact of gravity on the vacuum decay
rate~$\Gamma$. To this end, we consider the previous theory in a
curved spacetime background, including the Einstein--Hilbert term
governed by the Ricci scalar $R$. The Euclidean action now reads
\begin{equation}
\label{gaction}
S[\phi, g_{\mu\nu}]\ =\ \int \!d^4 x\, \sqrt{g} \:
\bigg(\! -\frac{1}{2}\, M^2_{\rm Pl}\, R\ +\ \frac{1}{2}\, g^{\mu \nu} 
\partial_\mu \phi \ \partial_\nu \phi\: +\: V(\phi) \bigg)\;,
\end{equation}
where $g_{\mu\nu}$ is the Euclidean spacetime metric, with
$g\equiv \det g_{\mu\nu}$, and $M_{\rm Pl}$ is the {\em reduced}
Planck mass. The latter is related to the ordinary Planck
mass~${M_{\rm P} \approx 1.9\times 10^{19}}$~GeV and the Newton's
constant $G_{\rm N}$ as follows:
\begin{equation}
\label{MPlanck}
M^2_{\rm Pl}\ \equiv\ \frac{M^2_{\rm P}}{8\pi}\ =\ \big(8\pi\, G_{\rm N}\big)^{-1}\; .
\end{equation}

Requiring now that the curved spacetime metric $g_{\mu\nu}$ be
$\mathbb{O}(4)$-symmetric, the (squared) line element\- may then be
expressed as
\begin{equation}
\label{metric}
ds^2\ =\ dr^2\: +\: \rho^2(r)\, d\Omega_3^2\;,
\end{equation}
where $d\Omega_3^2$ is the unit 3-sphere line element
and $\rho(r)$ is the volume radius of the 3-sphere at 
fixed $r$ coordinate~\cite{Coleman:1980aw}. The bounce configuration needed 
to calculate the false vacuum transition rate will be given now by
the field and the metric solutions, $\phi_{b}(r)$ and 
$\rho_{b}(r)$, to the coupled system of equations:
\begin{equation}
\label{gequation}
\ddot \phi\: +\: 3 \ \frac{\dot \rho}{\rho} \: \dot \phi\ 
=\ \frac{d V}{d \phi}\ , \qquad 
\dot \rho^2\ =\ 1\: +\: \frac{\kappa \rho^2}{3}
\bigg(\, \frac 1 2 \dot \phi^2\: -\: V(\phi) \bigg)\,,
\end{equation}
with $\kappa \equiv 8\pi G_{\rm N}$.  Note that the first equation
in~\eqref{gequation} replaces (\ref{meq}), while the second one is the
only non-trivial Einstein equation left by $\mathbb{O}(4)$ symmetry.
As before, we set~$V(\phi_{\rm fv})=0$ at the Minkowski false vacuum
$\phi_{\rm fv}$.  As we are interested in the decay of the false
vacuum to a true AdS vacuum, the appropriate boundary conditions read
\begin{equation} 
\label{bcond}
\phi_b(\infty)\ =\ \phi_{\rm fv}\;, \qquad \dot \phi_b(0)\ =\ 0\;, 
\qquad \rho_b(0)\ =\ 0\;.
\end{equation}
We must remark here that besides the non-trivial solution
$(\phi_b(r),\rho_b(r))$, the coupled\- system of
equations~\eqref{gequation} subject to \eqref{bcond} also admits a
trivial flat-space solution ${\phi(r) = \phi_{\rm fv}}$ with a
standard Euclidean metric $\rho(r)=r$. Moreover, the first
condition${}$ in~\eqref{bcond} implies  that as $r \to \infty$,
the bounce $\phi_b(r)$ approaches asymptotically the constant false
vacuum solution $\phi_{\rm fv}$, and  so $\rho_b(r)$ acquires the form of
a flat spacetime metric, since one has $\dot{\rho}_b(\infty) = 1$ by virtue of the
second equation in~\eqref{gequation}.

With the help of the Einstein equation,
$R_{\mu \nu}-\frac 1 2 g_{\mu \nu} R=\kappa T_{\mu \nu}$, where
$T_{\mu \nu} = \partial_\mu \phi\, \partial_\nu \phi - g_{\mu \nu} 
\left( \frac 1 2 \partial^\lambda \phi\, \partial_\lambda \phi +V(\phi) \right)$
is the energy-momentum tensor for the scalar field~$\phi$, the Ricci
scalar $R$ may then be easily determined by 
\begin{equation}
\label{trace}
\frac 1 \kappa \, R\ =\ g^{\mu\nu}\,\partial_\mu \phi \, \partial_\nu \phi\: +\: 4\, V(\phi) \; .
\end{equation}
Substituting (\ref{trace}) into (\ref{gaction}) and using the metric
(\ref{metric}) yields the simpler form for the action
\begin{equation}
\label{gravaction}
S[\phi,\rho]\ =\ -2 \pi^2 \int_0^\infty\! dr \, \rho^3\, V(\phi) \; .
\end{equation}
Evaluating the action (\ref{gravaction}) for the false vacuum
solution~$\phi_{\rm fv}$, we obtain
$S[\phi_{\rm fv},r] = 0$, leading to the tunnelling
exponent $B = S_b \equiv S[\phi_{b},\rho_{b}]$.

Both in the flat and curved spacetime cases, an 
important parameter is the size $\cal R$ of 
the bounce. This is defined as the value of $r$, for which
at $r = {\cal R}$ one has
\begin{equation}
\phi_b({\cal R})\ =\ \frac 1 2\, \phi_b(0)\; .
\end{equation}
By virtue of the bounce size~${\cal R}$, the prefactor~$D$
in~\eqref{gamma} may be estimated in the leading-order approximation.
In fact, to a good approximation, the lifetime $\tau=\Gamma^{-1}$ of
the false EW vacuum may be expressed in terms of~${\cal R}$ and the
age of Universe~$T_U$ as follows~\cite{Arnold:1991cv}, \cite{Isidori:2001bm,Branchina:2014rva}:
\begin{equation} 
\label{tau} 
\tau\ \simeq\ 
\frac{{\cal R}^4}{T_U^3}\; e^{B}\ =\ \frac{{\cal
		R}^4}{T_U^4}\ e^{S_b} \, T_U\;.  
\end{equation}
In all our numerical estimates, we will use~\eqref{tau} to compute
the false vacuum lifetime~$\tau$.

\subsection{Stability Analysis of the SM}

An application of~\eqref{tau} of immediate interest to us is the
computation of the lifetime of the EW vacuum of the SM effective
potential~$V_{\rm SM}(\phi)$. In this case, the field $\phi$ may be
identified with the gauge-invariant radial
direction~$\phi \equiv \sqrt{2}\,(\Phi^\dagger \Phi)^{1/2}$, where $\Phi$ is the
usual SM Higgs doublet. For field values $\phi \gg v$, the
Renormalization-Group (RG) improved effective potential may well be
approximated as
\begin{equation} 
\label{potential}
V_{\rm SM}(\phi)\ \approx\ \frac 1 4\, \lambda_{\rm SM} (\phi)\:\phi^4\,,
\end{equation}
where $\lambda_{\rm SM}(\phi)$ is the running quartic coupling
evaluated at the RG scale~$\mu =\phi$~\cite{Coleman:1973jx,Flores:1982rv}.

The running quartic coupling $\lambda_{\rm SM}(\mu )$ has been
determined by solving a system of RG equations up to three loop-level
accuracy~\cite{Rajantie:2016hkj}, after extending previous studies performed at two
loops~\cite{Degrassi:2012ry}. As noted in~\cite{Antipin:2013sga}, this may lead to
certain inconsistencies, when counting the number of loops in the beta
functions of the SM interactions, such as those for the Yukawa, gauge
and scalar quartic couplings. Since our purpose is to give an
order-of-magnitude estimate of the tunnelling time~$\tau$, we will not
use the proposed refinement in~\cite{Antipin:2013sga}, as it leads to minimal
changes, with no significant impact on the results of our
analysis. Instead, we will adopt a more simplified but equally robust
approach presented in~\cite{Burda:2016mou}.  In detail, we will approximate the
RG-improved SM effective potential~$V_{\rm SM}(\phi)$ up to two loops, by fitting the
scalar quartic coupling $\lambda_{\rm SM} (\phi )$ to a
three-parameter function:
\begin{equation}
\label{lam} 
\lambda_{\rm SM}(\phi)\ =\ \lambda_*\: +\: \alpha
\left ( \ln \frac{\phi}{M_{\rm P}} \right )^2\: +\: \beta \left ( \ln
\frac{\phi}{M_{\rm P}} \right )^4\,,
\end{equation}
with 
\begin{equation}
\label{p} 
\lambda_*\ =\ -0.013\;, \qquad 
\alpha\ =\ 1.4 \times 10^{-5}\,, \qquad 
\beta\ =\ 6.3 \times 10^{-8}\;.  
\end{equation} 
Our numerical evaluations will rely on the SM effective
potential~$V_{\rm SM}(\phi)$ in~\eqref{potential}, where
$\lambda_{\rm SM}(\phi)$ is given by (\ref{lam}) and (\ref{p}).

\begin{table}[t!]
	\renewcommand\arraystretch{1.5}
	\centering
	\begin{tabular}{|l|ccc|}
		\hline
		Background metric \qquad &	$\phi_0/M_{\rm P}$ \qquad & $V_0/M_{\rm P}^4$ \qquad &
		$\tau_{\rm SM}/T_U$\\ 
		\hline\hline
		Flat spacetime\qquad & $2.306$ \qquad & $-9.10\!\times\! 10^{-2}$ 
		\qquad & $10^{639}$ \\ \hline
		Curved  spacetime\qquad & $0.071$ \qquad & $-8.28\!\times\!10^{-8}$ \qquad &
		$10^{661}$ \\ 
		\hline
	\end{tabular}
	\caption{\em Numerical estimates of the lifetime~$\tau_{\rm SM}$
		of the SM EW vacuum in the absence of Planckian NP. The
		values $\phi_0 \equiv \phi_b (0)$ at the center of the bounce and 
		$V_0\equiv V_{\rm SM}(\phi_0)$ are also displayed.} 
	\label{tab:taboldres}
\end{table} 

Before we carry on studying the impact of NP on the stability of the EW
vacuum, we present in Table~\ref{tab:taboldres} the results of our
analysis of the tunnelling time~$\tau$ in the SM:
$\tau_{\rm SM}/T_U \sim 10^{639}$ (flat spacetime) and
~$\tau_{\rm SM}/T_U \sim 10^{661}$ (curved spacetime). These results
are in good agreement with those reported in the literature,
e.g.~see~\cite{Bentivegna:2017qry}.  For completeness, we include in the same table
the values of the profile $\phi(r)$ at the center of the bounce,
$\phi_0 \equiv \phi_b (0)$, as well as the value $V_0$ of the
potential~\eqref{potential} computed at $\phi_0$, i.e.~$V_0\equiv V_{\rm SM}(\phi_0)$.

\setcounter{equation}{0}
\section{Planckian New Physics Effects} 
\label{sec:PlanckNP}

In all our previous considerations, we have ignored the presence of
Planck-scale suppressed operators of the type $\phi^{2n}/M^{2n-4}$ in
the SM effective potential~$V_{\rm SM}(\phi)$ in~\eqref{potential},
where $M$ is of order~$M_{\rm P}$. Such operators could in principle
be generated by quantum gravity effects and as such, they cannot be
excluded {\it a priori} from~$V_{\rm SM}(\phi)$. If their contribution
to the SM potential happens to be negative, they may have a dramatic
destabilizing effect on our EW vacuum, as extensively discussed
in~\cite{Branchina:2013jra,Branchina:2014usa,Branchina:2014rva,Branchina:2015nda} for $n=3$. For reasons that will become
more clear below, we call such operators that contribute with a {\em
	negative} sign to~$V_{\rm SM}(\phi)$ as {\em harmful} operators.

Let us consider a set of distinct $\phi^{2n}$-models that could
effectively describe unknown Planckian NP effects. To this end, we
extend the SM effective potential as follows:
\begin{equation} 
\label{potV}
V_{2n}(\phi)\ =\ V_{\rm SM}(\phi)\: +\: V^{(2n)}_{\rm NP}(\phi)\; , 
\end{equation}  
where $n \geq 3$, and 
\begin{equation}
\label{new}
V^{(2n)}_{\rm NP}(\phi)\ =\ \frac{c_1}{2n} \ \frac{\phi^{2n}}{M^{2n-4}}\:
+\: \frac{c_2}{2(n+1)} \ \frac{\phi^{2(n+1)}}{M^{2n-2}}\ .
\end{equation}
Observe that all potentials $V_{2n}(\phi)$ in~\eqref{potV} reduce to
$V_{\rm SM}(\phi)$ for $\phi \ll M$, as its NP part,
$V^{(2n)}_{\rm NP}(\phi)$, becomes subdominant in this small-field
regime. For all the NP effective potentials~$V^{(2n)}_{\rm NP}(\phi)$, we
will assume that $c_1$ is negative, but $c_2$ is positive, so as to
ensure the convexity of the potential at high field values of
$\phi \gg M$. Thus, the first term~$\phi^{2n}/M^{2n-4}$ in~\eqref{new}
represents a harmful operator, which we will use from now on to characterize both
the $\phi^{2n}$-model and its effective potential~$V_{2n}(\phi)$.

As an interesting side remark, we note that the NP effective
potentials considered in~\eqref{new}, as well as the SUGRA effective
potentials that we will be studying in Section~\ref{sec:SUGRA},
satisfy to a fair degree the much recently discussed Weak Gravity
Conjecture (WGC)~\cite{Vafa:2005ui,ArkaniHamed:2006dz}.  According to
a refined version of the WGC~\cite{Obied:2018sgi,Grimm:2018ohb},
consistent effective theories of quantum gravity must satisfy two
basic criteria:
\begin{equation}
\label{eq:WGC}
\mbox{(i)}~~M_{\rm Pl}\, |\!|\mbox{\boldmath $\nabla$} V |\!|\ \ge\ c\, V\; ,\qquad
\mbox{(ii)}~~\Delta \phi\ \le\ d\, M_{\rm Pl}\; .
\end{equation}
Here, $V$ is the effective potential of the candidate theory,
$\mbox{\boldmath $\nabla$} V$ is its gradient in the field space,
$\Delta \phi \sim \phi$ refers to a domain of this field space, and
$c$ and $d$ are two positive constants of order 1, whose precise
values depend on the details of string compactification.  If the two
criteria in~\eqref{eq:WGC} are not met, the candidate theory is said
to belong to a kind of {\em swampland}, with no possible embedding
into a string theory. Rather interestingly, for all the examples
considered in the present study, the instanton dynamics, which is
crucial here for triggering EW vacuum decay
(cf.~Section~\ref{sec:thback}), requires a {\em negative} AdS
potential, for which $V(\phi ) \le 0$ and
$\phi \stackrel{<}{{}_\sim} M_{\rm P} \!=\! \sqrt{8\pi}\, M_{\rm Pl}$.
Hence, for this domain of the $\phi$-space, the first criterion~(i)
in~\eqref{eq:WGC} is trivially satisfied, whereas for the second
criterion~(ii) one only needs to tolerate values of $d$, as large as
$\sqrt{8\pi} \sim 5$. Consequently, for
$\phi \stackrel{>}{_\sim} M_{\rm P}$, the form of the effective
potentials $V^{(2n)}_{\rm NP}(\phi)$ needs to be strongly modified
according to the WGC criteria in~\eqref{eq:WGC}. However, such
modifications to the NP part $V^{(2n)}_{\rm NP}(\phi)$ of the SM
effective potential at the trans-Planckian regime have no significant
effect on the results of our analysis.

In the following, we will analyze numerically the impact of the
harmful Planck-scale NP operators~$\phi^{2n}/M^{2n-4}$ on the
tunnelling time~$\tau$ of the EW vacuum, for {\em all}~$n\ge 3$. To better
assess the relevance of these operators, we will simply set $c_1 = -2$
and $c_2 = 2$. Moreover, we will investigate two Planck-scale
scenarios with: (i)~$M=M_{\rm P}$ and (ii)~$M=M_{\rm P}/10$,  
for both a flat and a curved background metric.

\subsection{Planck-Scale Scenarios with {\boldmath $M = M_{\rm P}$}}
\label{sub:sub3.1}

\begin{figure}[t!] 
	\centering
	${}$\hspace{-9mm}\includegraphics[width=0.50\textwidth]{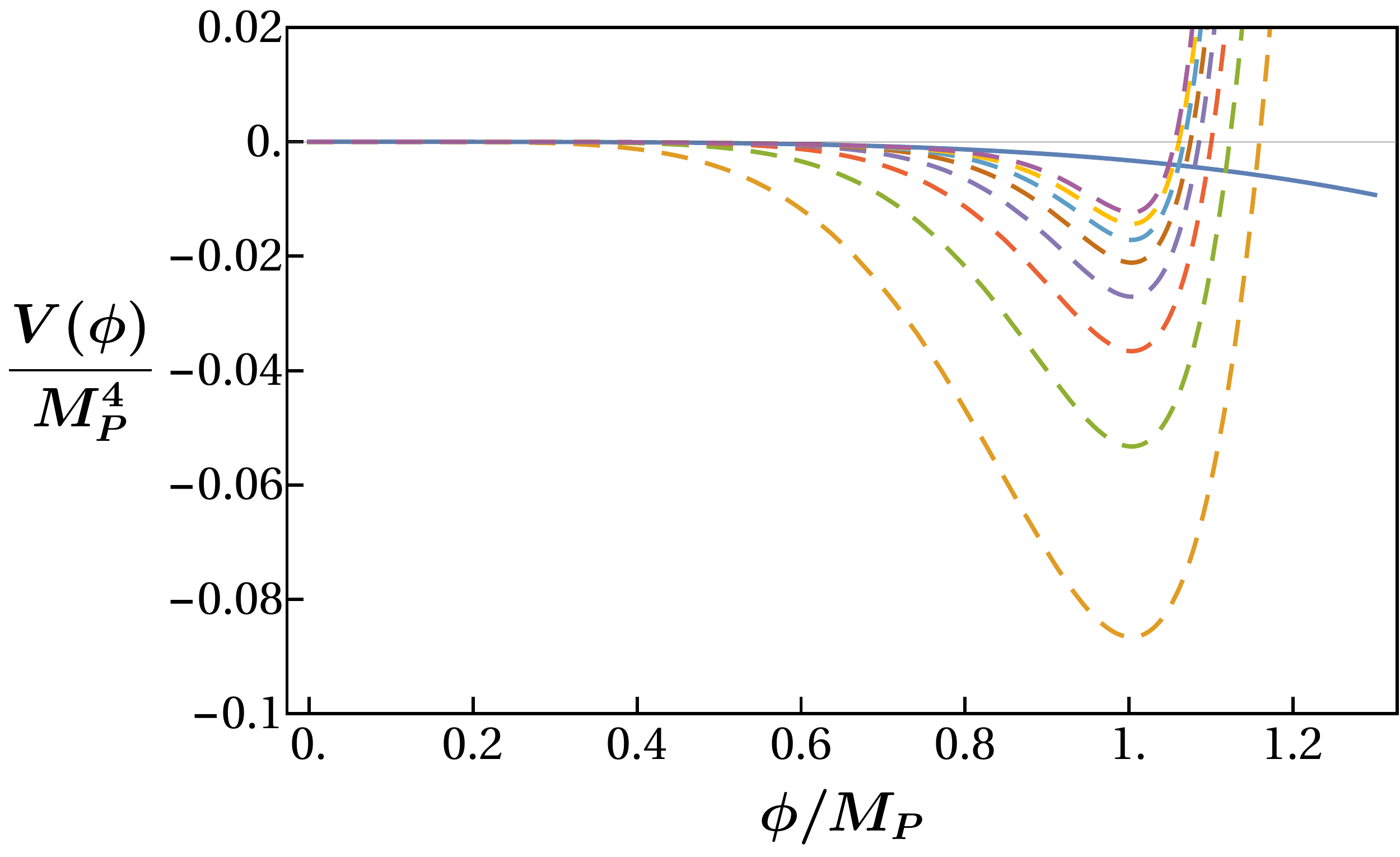}\\[7mm]
	\includegraphics[width=0.49\textwidth]{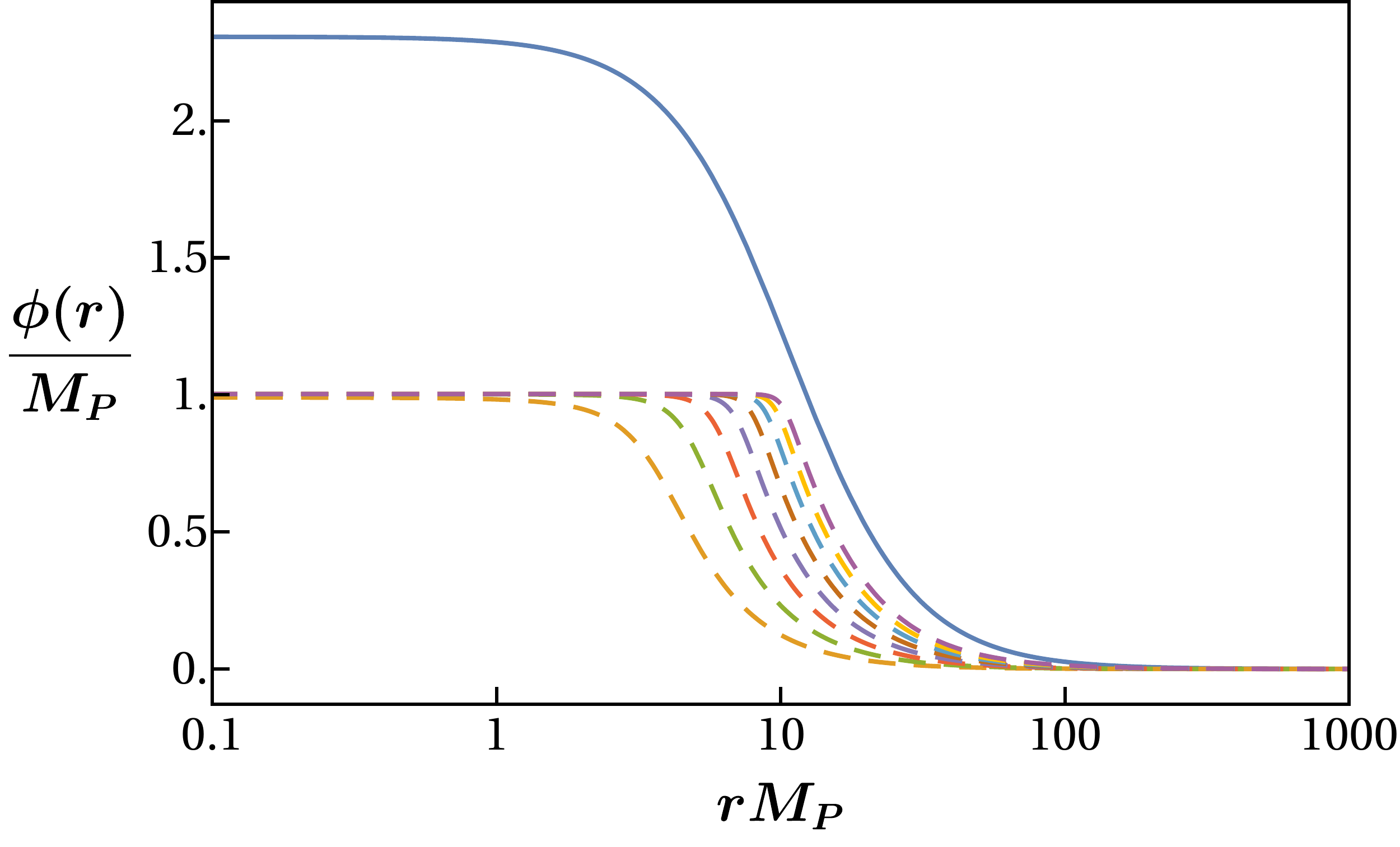}
	\caption{\em {\rm Upper panel}. Scalar potentials
		$V_{2n}(\phi)$ [cf.~\eqref{potV}], for
		$3\le n \le 10$, as functions of~$\phi$ (dashed
		lines), with $c_1\!=\! - c_2\!=\! -2$ and
		$M = M_{\rm P}$. The solid (blue) line corresponds to the SM
		potential~$V_{\rm SM}(\phi)$ given
		in~\eqref{potential}. {\rm Lower panel}. Radial
		profiles of bounce
		solutions~$\phi (r) \equiv \phi_b(r)$ (dashed lines)
		for the same class of Planck-scale scenarios
		evaluated for a flat spacetime metric. The solid
		(blue) line refers to the respective SM bounce.}
	\label{fig:pot}
\end{figure}

We first consider a class of $\phi^{2n}$-scenarios with scalar
potentials $V_{2n}(\phi)$ given by~\eqref{potV}, where the Planckian
NP scale $M$ is set equal to $M_{\rm P}$. As can be seen from the upper
panel of~Figure~\ref{fig:pot} and presented by dashed lines in multiple
colours, the negative contribution of the harmful Planck-scale
operators~$\phi^{2n}/M_{\rm P}^{2n-4}$ in~\eqref{new} produces a second
minimum in their respective effective potentials at
$\phi \sim M_{\rm P}$. For comparison, in the same panel we display
with a solid blue line the SM effective potential~$V_{\rm SM}(\phi)$
given in~\eqref{potential}.

A key quantity that determines the tunnelling decay time~$\tau$ of the EW
vacuum is the actual profile of the bounce
solutions~$\phi(r)\equiv \phi_b(r)$. These are depicted by dashed
lines in multiple colours on the lower panel of Figure~\ref{fig:pot} for a flat background
metric, where the solid line in blue corresponds to the SM bounce. Note that
all the bounces~$\phi(r)$ reach their highest value close to~$r = 0$,
thereby giving the largest support to the tunnelling exponent~$B$
in~\eqref{tau}. When normalising the effect of the harmful NP
operators to the one originating from the SM potential term
$\frac{1}4 \lambda \phi^4$, we get the ratio
\begin{equation}
\label{eq:ratio}
R_{2n}\ =\ \frac{2\, c_1}{n\, \lambda }\,\bigg(\frac{\phi(0)}{M_{\rm P}}\,\bigg)^{2n-4}\,.
\end{equation}
Since $\phi(0)/M_{\rm P} \sim 1$ for all $n\ge 3$, we expect that as
$n$ increases, $R_{2n}$ will decrease and the predictions for the EW
vacuum lifetime~$\tau$ will get closer to the SM value~$\tau_{\rm SM}$
presented in Table~\ref{tab:taboldres}. Indeed, this property is
observed in Table~\ref{tab:NPMP} for the flat spacetime case. In order
to get a lifetime~$\tau$ much larger than the age of the
Universe~$T_U$, we need to suppress all potentially harmful
operators~$\phi^{2n}/M_{\rm P}^{2n-4}$ up to $n = 6$, while one gets
$\tau \sim \tau_{\rm SM}$ when~$n \ge 50$.  In the next section, we
will outline a protective mechanism within a SUGRA framework, which
can in principle give rise to such a suppression.

\begin{table}[t!]
	\centering
	\renewcommand\arraystretch{1.5}
	\begin{tabular}{|c|lr||c|lr|}
		\hline
		& & & & & \\[-4mm]
		$\frac{\displaystyle \phi^{2n}}{\displaystyle M^{2n-4}}$ & $\tau/T_U$ & $\tau/T_U$~~ & 
		$\frac{\displaystyle \phi^{2n}}{\displaystyle M^{2n-4}}$ & $\tau/T_U$ & $\tau/T_U$~~~\\[-2mm] 
		$n$ & (flat) & (curved) & $n$ & (flat) & (curved) \\
		\hline\hline
		3 & $10^{-208}$ & $10^{-122}$ &                 7 & $10^{7}$   & $8.8 \!\times\! 10^{661}$ \\ 
		4 & $10^{-166}$ & $3.4\!\times\! 10^{661}$ & 8 & $10^{71}$  & $8.8 \!\times\! 10^{661}$\\ 
		5 & $10^{-114}$ & $8.8 \!\times\! 10^{661}$ & 9 & $10^{133}$ & $8.8 \!\times\! 10^{661}$\\ 
		6 & $10^{-55}$   & $8.8 \!\times\! 10^{661}$ & \hspace{-2mm}10& $10^{193}$
		& $8.8 \!\times\!10^{661}$\\
		\hline
	\end{tabular}
	\caption{\em Lifetime $\tau$ of the EW vacuum for a class of Planck-scale
		scenarios with harmful operators $\phi^{2n}/M^{2n-4}$, evaluated for
		a  flat and a curved background metric. As input values for the NP parameters, we set 
		$c_1\!=\! - c_2\!=\! -2$ and $M = M_{\rm P}$.
	}
	\label{tab:NPMP}
\end{table}

\begin{figure}[t!] 
	\centering
	${}$\hspace{-7mm}\includegraphics[width=0.50\textwidth]{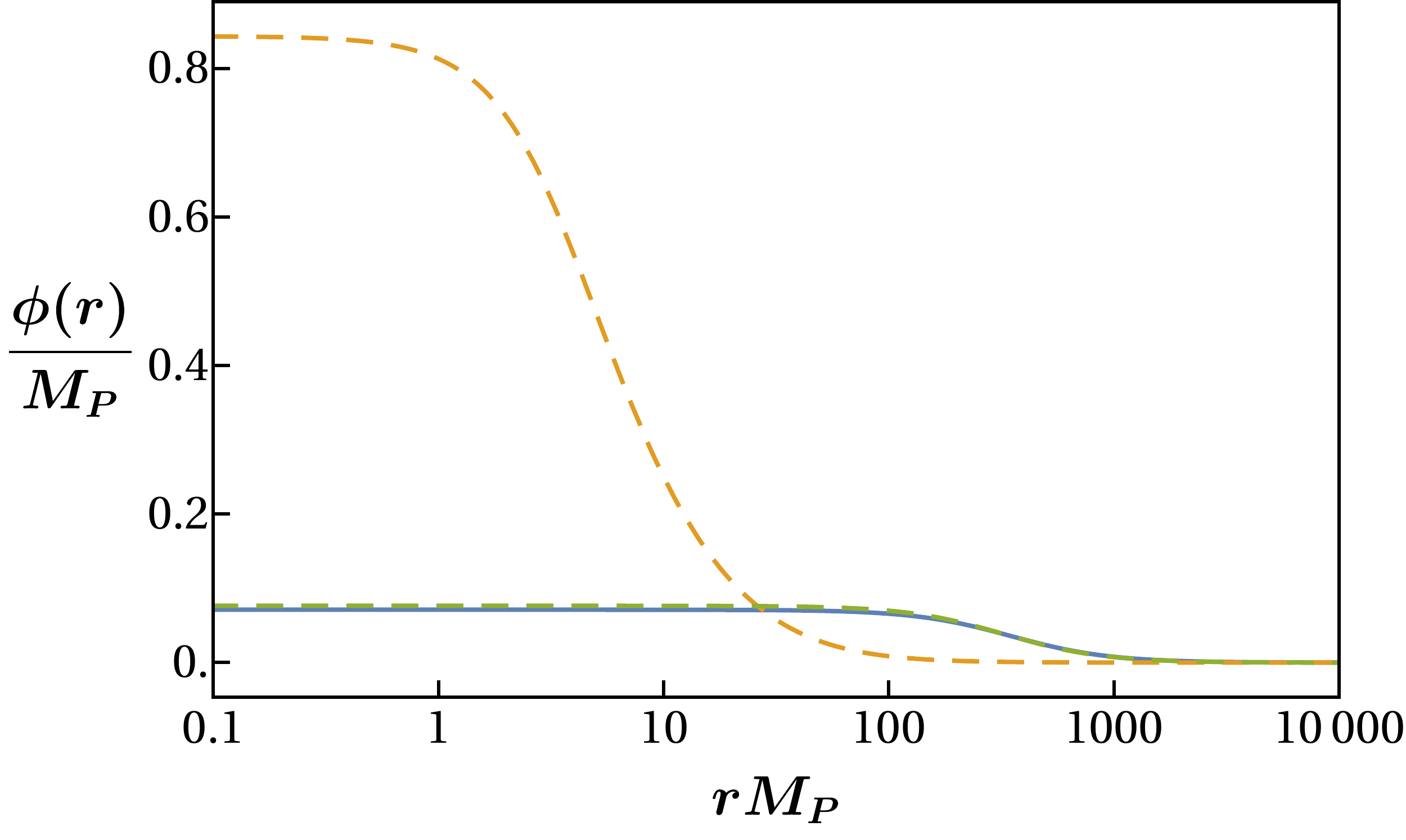}\\[7mm]
	\includegraphics[width=0.45\textwidth]{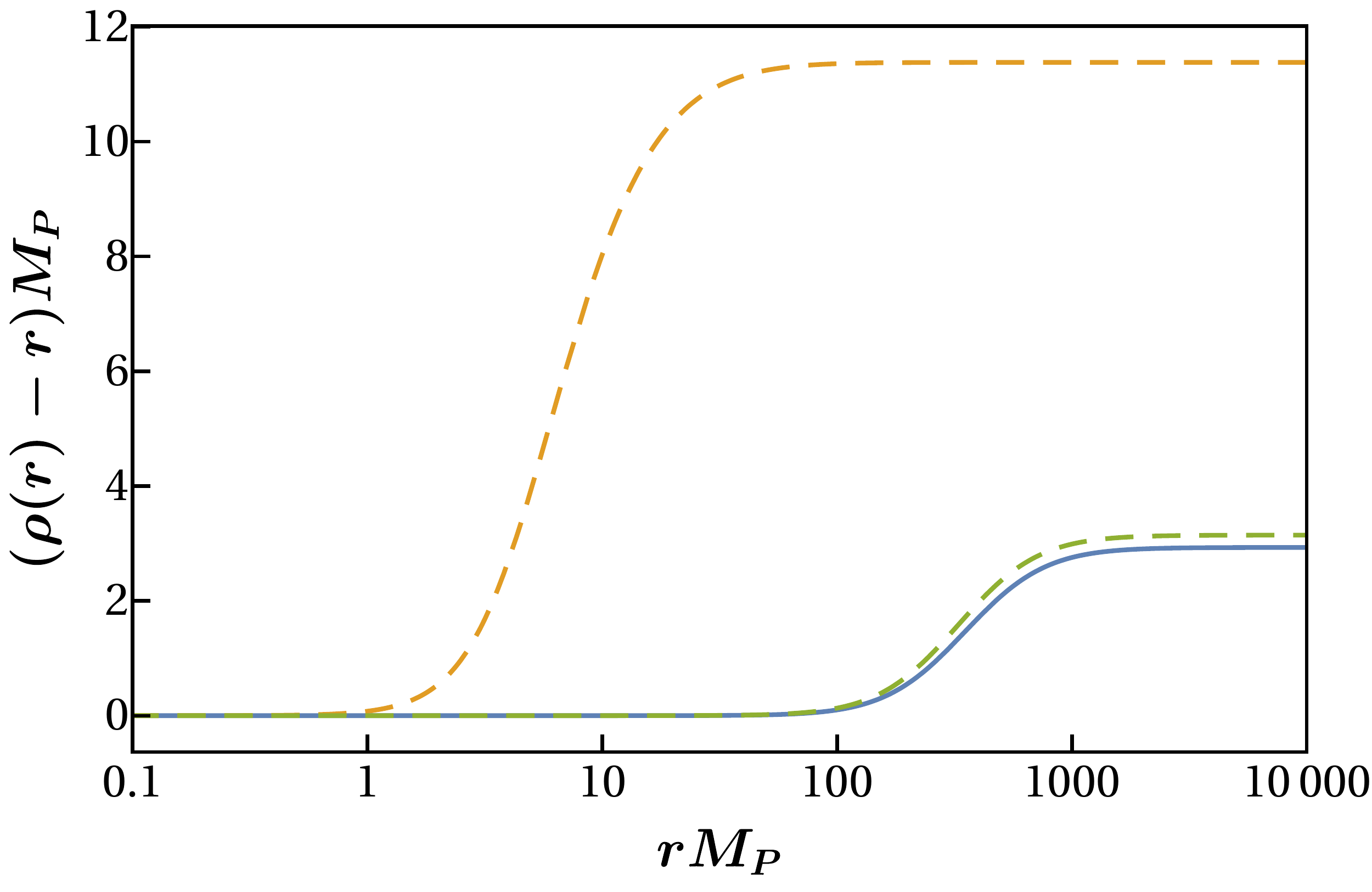}
	\caption{\em {\rm Upper panel}. Radial dependence of
		the bounce solutions~$\phi (r)$ (dashed lines) for
		the Planck-scale scenarios, with $n =3,\, 4$ and
		$M = M_{\rm P}$, evaluated for a curved spacetime
		metric. {\rm Lower panel}. Radial profiles for
		$\rho(r)-r$ (dashed lines) for the same scenarios
		and back\-ground metric.  The solid (blue) lines in
		the two panels show the bounce profiles in the SM. }
	\label{fig:PotCurvMP}
\end{figure}

Let us now investigate the effect of a curved background metric on the
EW vacuum\- lifetime $\tau$. Unlike in the flat spacetime, an important
novel aspect of the curved metric\- is that for increasing $n$, the
bounce solutions~$\phi (r)$ and $\rho (r)$ rapidly tend to the
corresponding SM bounces, as shown in Figure~\ref{fig:PotCurvMP}.  As
exhibited in~Table~\ref{tab:NPMP}, we obtain
$\tau \sim \tau_{\rm SM} \sim 10^{661}\,T_U$
(cf.~Table~\ref{tab:taboldres}), for all Planck-scale scenarios with
$n\ge 4$. This stabilizing effect of gravity on the EW vacuum may also
be attributed to the fact that for~${n\ge 4}$, one finds
$\phi (0) \sim 0.07$ which is smaller by more than one order of
magnitude from the corresponding value in the flat spacetime. As a
result, the size of NP contributions as represented by $R_{2n}$
in~\eqref{eq:ratio} will decrease more drastically as $n$ grows, for a
curved spacetime metric. In the next section, we will explore whether
this feature will persist for Planck-scale scenarios with a lower
quantum gravity scale~$M$.

\subsection{Planck-Scale Scenarios with {\boldmath $M = M_{\rm P}/10$}}

Proceeding as in the previous section, we will analyze a similar class
of $\phi^{2n}$-models, by assuming that the Planckian NP scale $M$ is
now one order of magnitude smaller, i.e.~$M = M_{\rm P}/10$. Such a
choice may be motivated by the fact that the relevant energy\- scale of
quantum gravity that enters Einstein's equation is the reduced Planck
mass~$M_{\rm Pl}$~[cf.~\eqref{MPlanck}], rather than the ordinary
Planck mass~$M_{\rm P}$.

\begin{figure}[t!] 
	\centering
	${}$\hspace{-16mm}\includegraphics[width=0.50\textwidth]{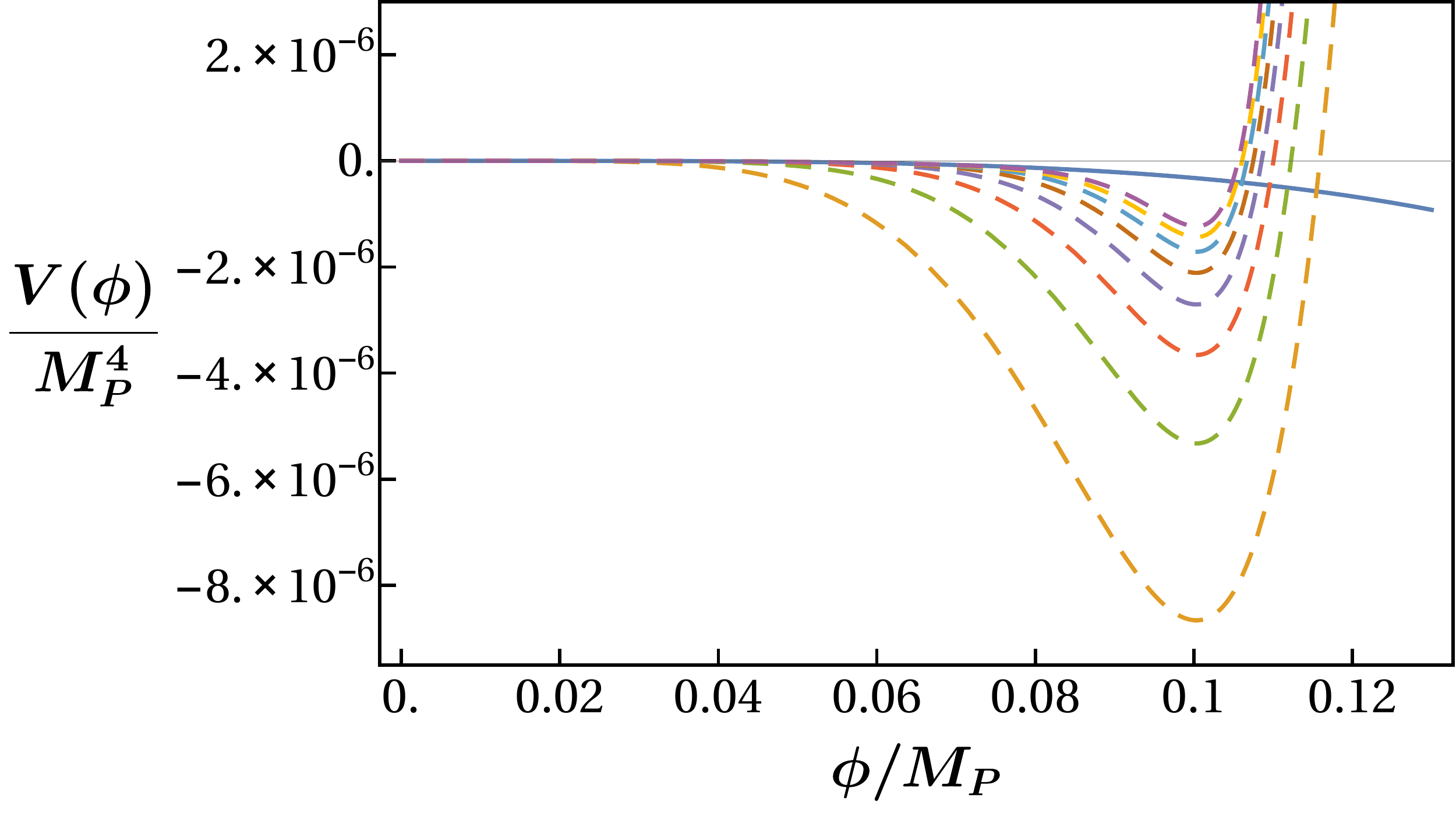}\\[10mm]
	\includegraphics[width=0.49\textwidth]{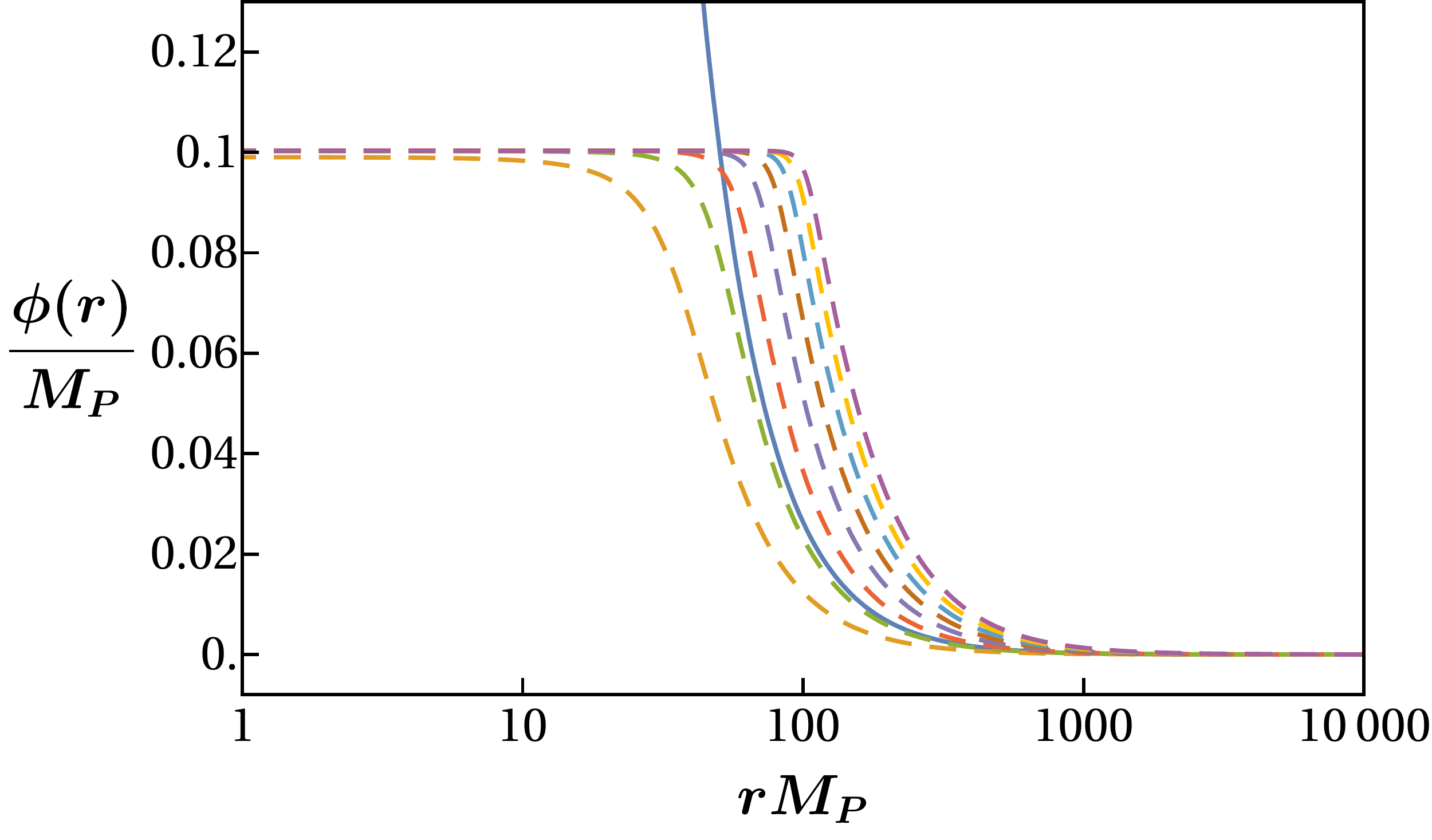}
	\caption{\em The same as in Figure~\ref{fig:pot}, but setting instead $M = M_{\rm P}/10$.}
	\label{fig:PotPhiFlat}
\end{figure}

\begin{table}[t!]
	\centering
	\renewcommand\arraystretch{1.5}
	\begin{tabular}{|c|ll||c|ll|}
		\hline
		& & & & & \\[-4mm]
		$\frac{\displaystyle \phi^{2n}}{\displaystyle M^{2n-4}}$ & $\tau/T_U$ & ~~$\tau/T_U$ & 
		$\frac{\displaystyle \phi^{2n}}{\displaystyle M^{2n-4}}$ & $\tau/T_U$ & ~~$\tau/T_U$\\[-2mm] 
		$n$ & (flat) & (curved) & $n$ & (flat) & (curved) \\
		\hline\hline
		3 & $10^{-204}$ & $10^{-203}$ & 7 & $10^{12}$  & $10^{21}$\\ 
		4 & $10^{-162}$ & $10^{-160}$ & 8 & $10^{76}$  & $10^{87}$\\ 
		5 & $10^{-110}$ & $10^{-106}$ & 9 & $10^{138}$ & $10^{152}$\\ 
		6 & $10^{-51}$ &  $10^{-44}$ & \hspace{-2mm}10 & $10^{198}$ & $10^{214}$\\
		\hline
	\end{tabular}
	\caption{\em Lifetime $\tau$ of the EW vacuum for a class of Planck-scale
		scenarios that include harmful operators $\phi^{2n}/M^{2n-4}$, with $M =
		M_{\rm P}/10$, evaluated for both 
		a  flat and a curved background metric. 
	}
	\label{tab:NPMP10}
\end{table}

From the upper panel of Figure~\ref{fig:PotPhiFlat}, we observe that
the minimum of the potentials~$V_{2n}$ is now located to a smaller
value at $\phi_{\rm min} \sim M_{\rm P}/10$.  The profiles of the
bounces~$\phi (r)$ for a flat spacetime metric are presented by dashed
lines in various colours in the lower panel of
Figure~\ref{fig:PotPhiFlat}, while the solid (blue) line stands for
the SM bounce. In order to assess the impact of gravity, we give in
Table~\ref{tab:NPMP10} the lifetime~$\tau$ of the EW vacuum, for both a
flat and a curved spacetime background. As opposed to the previous
scenarios, we now observe that the impact of gravity is less 
significant, and restoration of the SM prediction for~$\tau$ takes place
for $n \ge 41$. 

Comparing the flat-spacetime results exhibited in 
Tables~\ref{tab:NPMP10} and~\ref{tab:NPMP}, we notice that the
predicted values for the tunnelling times~$\tau$ for each $n$ turn out
to be close. As before, we may understand this result by looking
at the ratios,
\begin{equation}
\label{eq:ratio2}
R_{2n}\ =\ \frac{2\,c_1}{n\, \lambda }\,\bigg(\frac{\phi
	(0)}{M}\bigg)^{2n-4} =\ 
\frac{2\,c_1}{n\, \lambda }\,\bigg(
\frac{10\,\phi (0)}{M_{\rm P}}\bigg)^{2n-4}\,.
\end{equation}
Unlike the previous case $M=M_{\rm P}$, an extra factor $10^{2n-4}$
now appears, because we have~$M=M_{\rm P}/10$.  As can be seen from
the lower panel of Figure~\ref{fig:PotPhiFlat}, the maximum of all
bounces reached at their origin ($r=0$) approaches the value:
$\phi(0)/M_{\rm P} \sim 0.1$. Hence, the enhancement
factor~$10^{2n-4}$ in~\eqref{eq:ratio2} gets compensated by a
corresponding factor
$\big(\phi(0)/M_{\rm P}\big)^{2n-4}\sim 10^{-(2n - 4)}$. 
As a consequence of this cancellation, the order-of-magnitude
estimates of the tunnelling time~$\tau$ for the two Planck-scale
scenarios, with $M = M_{\rm P}$ and $M = M_{\rm P}/10$, will be comparable.

\begin{figure}[t!] 
	\centering
	${}$\hspace{-13mm}\includegraphics[width=0.50\textwidth]{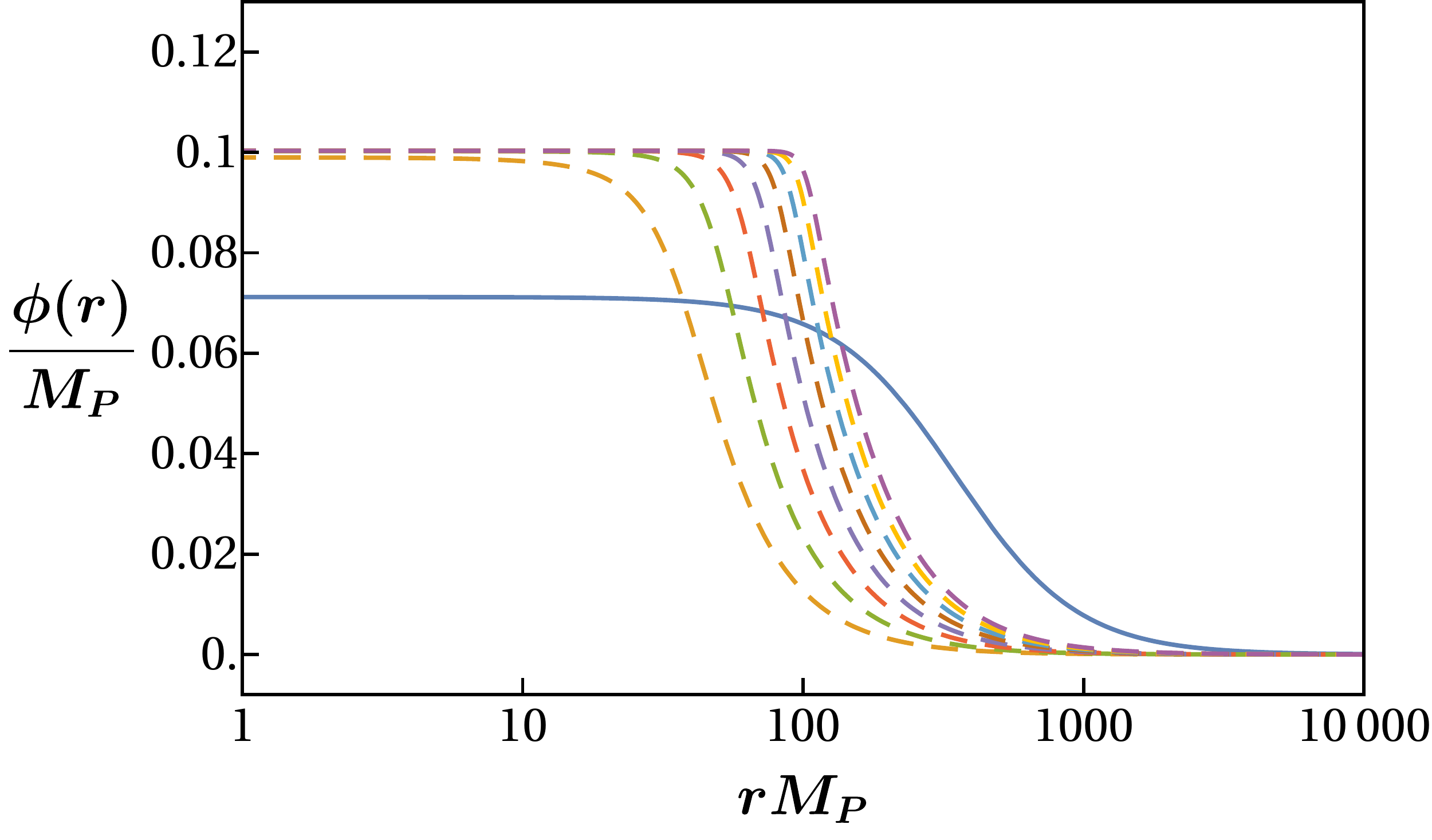}\\[7mm]
	\includegraphics[width=0.45\textwidth]{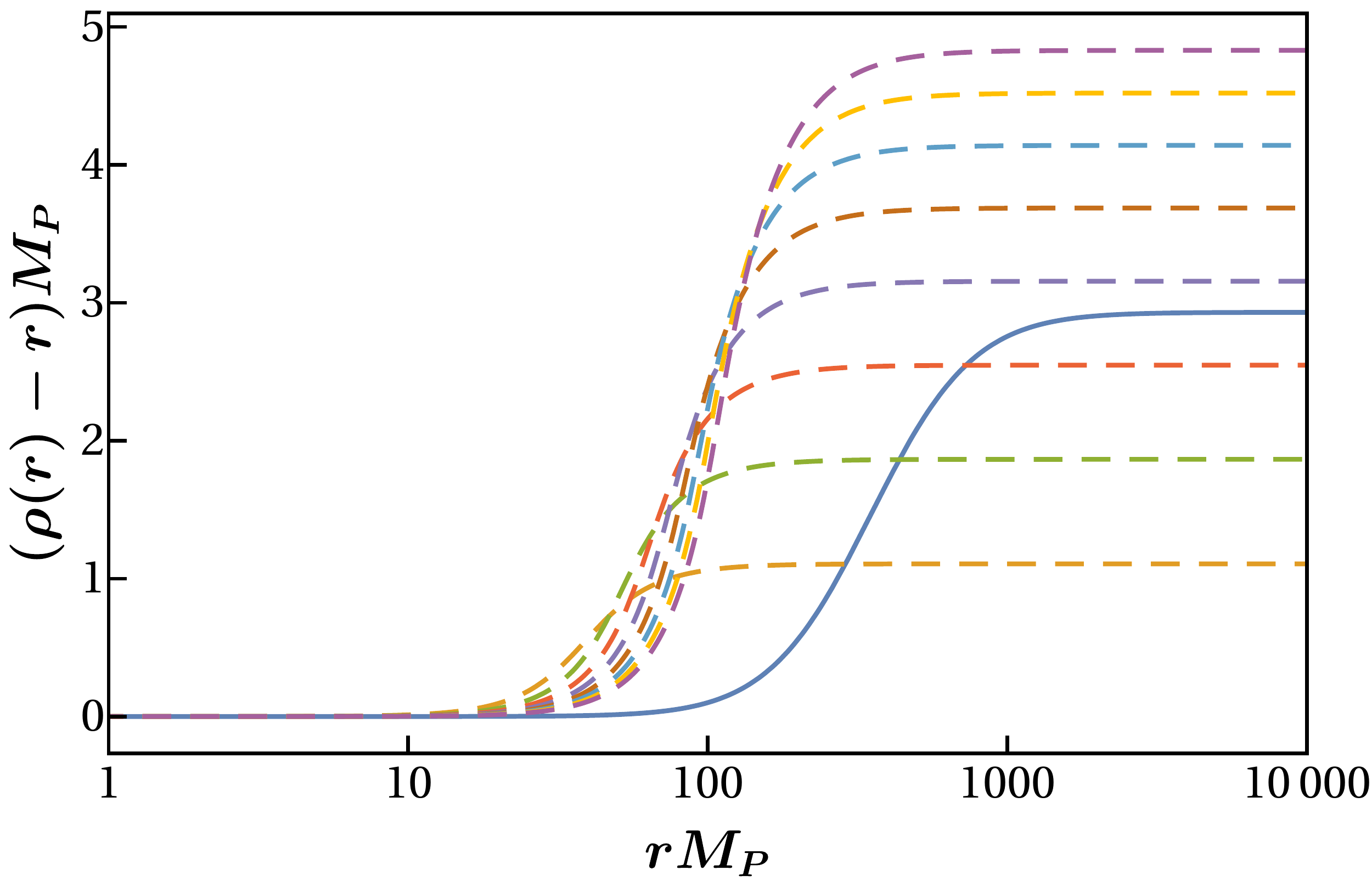}
	\caption{\em The same as in Figure~\ref{fig:PotCurvMP}, for 
		scenarios with $3\le n \le 10$ and $M =
		M_{\rm P}/10$.}
	\label{fig:PotCurvMP10}
\end{figure}

Let us now turn our attention to the curved spacetime analysis and the
numerical estimates of the EW vacuum lifetime~$\tau$ given in
Table~\ref{tab:NPMP10}. As mentioned earlier, the impact of gravity
on~$\tau$ is minimal in this case. This can be better understood by
analysing the profile for the bounce solutions~$\phi (r)$
and~$\rho (r)$, for $n \ge 3$. As shown in
Figure~\ref{fig:PotCurvMP10}, the bounces~$\phi(r)$ quickly approach
the ones found above in Figure~\ref{fig:PotPhiFlat} (lower panel) for
the flat spacetime metric. Hence, we expect for the EW vacuum
lifetime~$\tau$ to be less affected by the presence of gravity,
becoming independent of the radial coordinate~$\rho (r)$.

The above exercise illustrates how the occurrence of harmful operators
in Planckian NP theories that happen to realise a relatively low scale
of quantum gravity~$M$ face a serious destabili\-zation problem of the
EW vacuum.  In the next section, we will discuss mechanisms that can
naturally suppress the presence of leading harmful operators to
sufficiently higher powers of~$n$, within a minimal SUGRA framework.

\setcounter{equation}{0}
\section{Protective Mechanisms in SUGRA}
\label{sec:SUGRA}

Given our ignorance of a realistic UV-complete theory of quantum gravity,
Planck-scale gravi\-tational effects are usually treated within the
context of an effective field theory by considering {\em all}
possible gauge-invariant non-renormalizable operators suppressed by
inverse powers of a high-scale mass~$M$, which is typically of the
order of the {\em reduced} Planck mass~$M_{\rm Pl} \approx 2.4\times 10^{18}$~GeV.
Specifically, gravitational effects on the SM scalar
potential${}$~$V_{\rm SM}(\phi )$ along the gauge-invariant field
direction~$\phi = \sqrt{2}\,(\Phi^\dagger \Phi)^{1/2} \ge 0$, where $\Phi$
is the SM Higgs doublet, may be described by the effective potential
\begin{equation}
\label{eq:Vpot}
V(\phi )\ =\ V_{\rm SM}(\phi)\: +\: \sum\limits_{n = 3}^\infty\, \frac{\lambda_{2n}}{2n}\,
\frac{\phi^{2n}}{M^{2n - 4}}\ ,
\end{equation}
with $V_{\rm SM}(\phi) = -\,m^2\phi^2/2\: +\: \lambda \phi^4/4$.
Depending on the sign and size of the coefficient~$\lambda_6$ for the
leading Planck-scale suppressed operator $\phi^6/M^2$, the lifetime of
the EW vacuum can vary by many orders of
magnitude~\cite{Branchina:2013jra}. In particular, if $\lambda_6$ is
negative and $|\lambda_6 |$ large, the operator $\phi^6/M^2$ then becomes {\em
	harmful} and could lead to a dramatic destabilization of the EW
vacuum, for both flat and curved spacetime
backgrounds~\cite{Bentivegna:2017qry}.  In the following, we will show
how SUGRA embeddings of the SM~\cite{Nilles:1983ge} could protect the
EW vacuum from rapid decay up to very large values of the soft
SUSY-breaking scale~$M_{\cal S}$, above the so-called SM
metastability scale of~$10^{11}$~GeV.

To start with, let us first consider the Minimal Supersymmetric extension of
the Standard${}$ Model (MSSM), in which only Planck-mass suppressed
non-renormalizable operators involving the Higgs chiral superfields
$\widehat{H}_{1,2}$ are considered. In other words, we ignore for
simplicity non-renormalizable operators of all other chiral
superfields in the effective superpotential ${\cal \widehat{W}}$. In a
SUGRA framework, ${\cal \widehat{W}}$ will then be given by
\begin{equation}
\label{eq:Weff}
{\cal \widehat{W}}\ =\ {\cal \widehat{W}}_0\: +\: \mu\, \widehat{H}_1
\widehat{H}_2\: +\: \sum\limits_{n = 2}^\infty\, \frac{\rho_{2n}}{2n}\,
\frac{(\widehat{H}_1\widehat{H}_2)^n}{M^{2n - 3}}\ ,
\end{equation}
where 
\begin{equation}
\label{eq:WSM}
{\cal \widehat{W}}_0\ = \ h_l \,\widehat{H}_1\widehat{L}\widehat{E}\: +\: h_d
\,\widehat{H}_1\widehat{Q}\widehat{D}\: +\:  h_u \,\widehat{H}_2\widehat{Q}\widehat{U}
\end{equation}
is the usual MSSM superpotential without the $\mu$ term, and
$\widehat{H}_1$, $\widehat{H}_2$ are the chiral superfields for the
two Higgs doublets, $\widehat{Q}$, $\widehat{L}$ are the chiral
superfields for the quark and lepton left-handed iso-doublets, and
$\widehat{U}$, $\widehat{D}$, $\widehat{E}$ are their respective
right-handed iso-singlet counterparts. Here we follow the conventions
of~\cite{Carena:2000yi}. A comprehensive review of the EW sector of
the MSSM is given in~\cite{Djouadi:2005gj}. Note that in
writing~\eqref{eq:WSM}, we have suppressed all flavour indices from
the lepton and quark Yukawa couplings $h_l$, $h_d$ and~$h_u$.

To simplify matters, we consider that all our SUGRA embeddings
are based on a minimal Kaehler potential ${\cal \widehat{K}}$ given by
\begin{equation}
\label{eq:Kmin}
{\cal \widehat{K}}\ \equiv\ {\cal K} (\widehat{\varphi}_i^*\,,\,
\widehat{\varphi}_i)\ =\ \widehat{H}^\dagger_1 \widehat{H}_1\ +\
\widehat{H}^\dagger_2 \widehat{H}_2\ +\ \dots\; ,
\end{equation}
where $\widehat{\varphi}_i$ is a generic chiral superfield and
contributions from SU(2)$_L$ and U(1)$_Y$ vector superfields are not
shown.  At the tree level, the scalar SUGRA potential $V$ may be
written as a sum of three terms: $V = V_F + V_D + V_{\rm br}$, since
it receives three contributions from: (i) $F$-terms ($V_F$),
(ii)~$D$-terms~($V_D$), and (iii) the so-called SUSY-breaking
terms~($V_{\rm br}$) induced by spontaneous breakdown of SUGRA that
may occur in the so-called hidden sector of the
theory~\cite{Nilles:1983ge}. In particular, the $F$- and $D$-terms of
the potential $V$ may be calculated from the general expressions:
\begin{eqnarray}
\label{eq:VF}
\nonumber V_F & = & e^{{\cal K}/M^2_{\rm Pl}}\ \bigg[ \bigg( {\cal W}_{,\, i} \: +\:
\frac{{\cal K}_{,\, i}}{M^2_{\rm Pl}}\, {\cal W}\bigg)\, G^{-1,\,
	i\bar{j}}\\ 
& &\times\, \bigg( {\cal W}_{,\, \bar{j}} \: +\:
\frac{{\cal K}_{,\, \bar{j}}}{M^2_{\rm Pl}}\, {\cal W}^*\bigg)\
-\ 3\, \frac{|{\cal W}|^2}{M^2_{\rm Pl}}\, \bigg]\ ,\\
\label{eq:VD}
V_D & = & \frac{g^2}{2} f^{-1}_{ab} D^a D^b\;,  
\end{eqnarray}
where ${\cal W} \equiv {\cal W}(\varphi_i)$,
${\cal K} \equiv {\cal K}(\varphi_i^* , \varphi_i)$,
${\cal W}_{,\, i} \equiv \partial {\cal W}/\partial \varphi_i$,
${\cal K}_{,\, i} \equiv \partial {\cal K}/\partial \varphi_i$,
${\cal K}_{,\, \bar{i}} \equiv {\cal K}^*_{,\, i}$ etc, for a generic
scalar field $\varphi_i$, and $G^{-1,i\bar{j}}$ is the inverse of the
Kaehler-manifold metric:
$G_{i\bar{j}} = {\cal K}_{,\,i\bar{j}} = \partial^2 {\cal K}/(\partial
\varphi_i \partial \varphi^*_j)$.
In addition, $g$ is a generic gauge coupling, e.g.~of SU(2)$_L$, $f_{ab}$
is the gauge kinetic function taken to be minimal,
i.e.~$f_{ab} = \delta_{ab}$, and
$D^a = {\cal K}_{,\, \varphi} T^a \varphi$ are the so-called
$D$-terms, where $T^a$ are the generators of the gauge group. Finally,
the SUSY-breaking Higgs potential $V^H_{\rm br}$ generated from the
effective superpotential in~\eqref{eq:Weff} is given by
\begin{eqnarray}
\label{eq:Vbr}
\nonumber V^H_{\rm br} &=& m^2_1\,|H_1|^2 + m^2_2\,|H_2|^2\\
&&+\: \bigg( B \mu\, H_1H_2 + \sum\limits_{n = 2}^\infty\, A_{2n}\,
\frac{(H_1 H_2)^n}{M^{2n - 3}} + {\rm H.c.}\bigg)\; .\qquad\quad
\end{eqnarray}
For the purpose of this study, we will assume that the $\mu$-term and
the soft mass parameters${}$~$m^2_{1,2}$ and $B\mu$ are of order
$M_{\cal S}$, but all other SUSY-breaking $A$-terms~$A_{2n}$ could be as
large as $M$.  Although such an unusual assumption does not sizeably
destabilize\- the gauge hierarchy for
$M_{\cal S} \stackrel{<}{{}_\sim} 10$~TeV, it can still significantly
affect the predictions for the Higgs-boson mass spectrum.  Here, we
will not address the mechanism\- causing this large hierarchy between
the soft parameters and the higher order~$A$-terms,
as it strongly depends on the details of the hidden sector in which
SUSY gets spontaneously broken~\cite{Nilles:1983ge}.

Let us for the moment consider the SUSY limit of the MSSM, by ignoring
the induced SUSY-breaking terms $V^H_{\rm br}$ in the scalar
potential~$V$.  Assuming that the $\mu$-term is of order $M_{\cal S}$
and so negligible when compared to $M_{\rm Pl}$, the renormalizable
part of the MSSM potential, denoted by~$V_0$, has an $F$- and $D$-flat
direction associated with the gauge-invariant operator
$\widehat{H}_1 \widehat{H}_2$. In the absence of the $\mu$-term, the
scalar field configuration:
\begin{equation}
\label{eq:Dflat}
H_1\ =\ \frac{1}{\sqrt{2}}\left( \begin{array}{c} 
\phi\\ 
0\end{array} \right)\;,\qquad 
H_2\ =\  \frac{e^{i\xi}}{\sqrt{2}}\, \left( \begin{array}{c} 
0\\
\phi \end{array} \right)\; ,
\end{equation}
with $\xi \in [0,2\pi)$ and all other scalar fields taken at the
origin, gives rise to an exact flat direction for~$V_0$,
i.e.~$\partial V_0/\partial \phi = 0$. Here $\phi$ is a {\em positive}
scalar field background with canonical kinetic term that parameterizes
the $D$-flat direction. The CP-odd angle $\xi$ indicates that the flat
directions for $H_1$ and $H_2$ may also differ by an arbitrary
relative phase~$\xi$.  Hence, the parameters $(\phi ,\xi )$ describe
fully the $D$-flat direction of interest.  Now, in the flat-space
limit $M_{\rm Pl} \to \infty$, $V_{0F}$ is positive, implying that
$V_0 = V_{0F} + V_{0D} \ge 0$, where the equality sign holds along a
flat direction, such as the $\phi$-direction.

The above property of a non-negative potential will generically
persist in the minimal${}$ SUGRA for the full observable-sector
potential $V$, namely upon the inclusion of gauge-invariant
non-renormalizable operators consisting only of MSSM fields.  To see
this, we first write the MSSM superpotential~${\cal \widehat{W}}$ as
the sum: ${\cal \widehat{W}} = \sum_a {\cal \widehat{W}}^a$, where
${\cal \widehat{W}}^a$ is an arbitrary superpotential term labelled by
the index $a$. Then, we notice that the only negative contribution
to~$V$ can potentially come from both the cross terms and the last
term that occur in $V_F$ given in~\eqref{eq:VF}. In particular, up to
an overall positive factor $e^{{\cal K}/M^2_{\rm Pl}}$, we have
\begin{eqnarray}
\label{eq:VFcross}
\nonumber V_F &\supset&  \sum_{a,b}\,\bigg[\bigg( \frac{{\cal
		K}_{,\,\bar{i}}}{M^2_{\rm Pl}}\,  {\cal W}^a_{,\, i}\, {\cal
	W}^{b\,*}\: +\: {\rm H.c.}\,\bigg)\: -\:  3\, \frac{{\cal
		W}^a{\cal W}^{b *}}{M^2_{\rm Pl}}\bigg] \\
	&&=\ \sum_{a,b} 
\Big( N_a + N_b - 3\Big)\, \frac{{\cal W}^a{\cal W}^{b\,*}}{M^2_{\rm Pl}}\; .
\end{eqnarray}
In arriving at the last equality in~\eqref{eq:VFcross}, we used the
fact that ${\cal K}_{,\,\bar{i}} {\cal W}^a_{,\, i} = N_a {\cal W}^a$
in minimal SUGRA, where $N_a$ is the number of scalar fields
$\varphi_i$ present in ${\cal \widehat{W}}^a$.  Given that
$N_{a,b} \ge 2$ for all superpotential terms in the MSSM, the last
expression on the RHS of~\eqref{eq:VFcross} will be non-negative, with
a possible exception specific field configurations for which
$N_a \neq N_b$. Thus, barring fine-tuning, the $F$-term potential
$V_F$ of the observable sector in~\eqref{eq:VF} will be
non-negative~$^{2}$\footnotetext[2]{Instead, hidden-sector chiral superfields
	$\widehat{Z}$ can lead to a negative contribution to $V_F$ via
	SUSY-breaking effects from a Polonyi-type superpotential
	${\cal \widehat{W}}_{\rm hidden} = m^2 ( \widehat{Z} + \beta )$, for which
	${N_{a,b} \le 1}$. This negative contribution is even desirable, as it can
	be used to fine-tune the cosmological constant to its observed small
	value. Similarly, non-minimal Kaehler potential may also lead to
	negative contributions to~$V_F$. A more detailed discussion is given
	in~\cite{Nilles:1983ge}.}.
Since $V_D\ge 0$ as well, the complete MSSM scalar potential,
including the infinite series of the non-renormalizable operators,
will generically be non-negative.

The above situation changes drastically, if the SUSY-breaking
$A$-terms as given in~\eqref{eq:Vbr} are added to the scalar
potential $V$. For illustration, let us consider the minimally
extended MSSM superpotential
\begin{equation}
\label{eq:Wrho4}
{\cal \widehat{W}}\ =\ {\cal \widehat{W}}_0\: +\: \mu\, \widehat{H}_1
\widehat{H}_2\: +\: \frac{\rho_4}{4}\,
\frac{(\widehat{H}_1\widehat{H}_2)^2}{M}\ ,
\end{equation}
which induces the SUSY-breaking potential
\begin{eqnarray}
\label{eq:Vbr4}
\nonumber V^H_{4,{\rm br}} &=& m^2_1\,|H_1|^2 + m^2_2\,|H_2|^2\\
&& +\: \bigg( B \mu\, H_1H_2 + A_4\,
\frac{(H_1 H_2)^2}{M}\ +\ {\rm H.c.}\bigg),\qquad\quad
\end{eqnarray}
for the Higgs sector. For simplicity, we envisage a scenario for which
$m^2_{1,2} \ll B\mu$.  Moving\- along the $D$-flat direction
as stated in~\eqref{eq:Dflat}, and upon ignoring radiative corrections\- for field
values $\phi > M_{\cal S}$, the leading part of the scalar potential takes on the
simple form:
\begin{eqnarray}
\label{eq:VHiggs4}
\nonumber V_4(\phi) &=& e^{\phi^2/M^2_{\rm Pl}}\, \bigg[ -\, \frac{m^2}{2}\,\phi^2\: +\: 
\frac{{\rm Re}(e^{2i\xi}A_4)}{2M}\, \phi^4\\ &&+\: \frac{|\rho_4|^2}{8}\,
\frac{\phi^6}{M^2}\, \bigg( 1 + \frac{5}{32}\,\frac{\phi^2}{M^2_{\rm Pl}} + 
\frac{1}{32}\,\frac{\phi^4}{M^4_{\rm Pl}} \bigg)\bigg]\, ,\qquad\qquad
\end{eqnarray}
where higher-order terms proportional to $|\mu |/M
\stackrel{<}{{}_\sim} M_{\cal S}/M \ll 1$
were neglected and $m^2 = |e^{i\xi}\,B\mu - |\mu|^2|$ is arranged\- to be of the
required EW order.  Note that even if $A_4 > 0$, the field direction~\eqref{eq:Dflat}
with $\xi = \pi/2$ will make the coefficient ${\rm Re}(e^{2i\xi}A_4)$
entering the potential $V_4$ in~\eqref{eq:VHiggs4} negative. If $A_4$
is comparable to $M$, the quartic coupling
$\phi^4$ can become both sizeable and negative, giving rise to a
potential $V_4$ that develops a new minimum of order
$M/|\rho_4|$, far away from its SM value. On the other hand, the
higher powers $\phi^6$, $\phi^8$ and $\phi^{10}$ are all proportional to
the positive coefficient $|\rho_4|^2$, thereby ensuring the convexity
of the potential $V_4$. Clearly, this exercise shows that SUSY
is rather effective in protecting the stability of the EW vacuum from
{\em unknown} Planck-scale gravitational effects, unless the induced
SUSY-breaking coupling $A_4$ happens to be of order $M \sim M_{\rm Pl}$.

In the following, we wish to argue that SUSY may still be effective
for protecting the stability of our EW vacuum, even for extreme
scenarios with $A_{2n} \sim M_{\rm Pl}$ and $M_{\cal S}$ above
the metastability scale of order~$10^{11}$~GeV, along the lines of 
split-SUSY~\cite{ArkaniHamed:2004fb,Giudice:2004tc}. 
To this end, let us
first consider the following discrete symmetry transformations on the
chiral superfields:
\begin{eqnarray}
\label{eq:Rsyms}
\Big(\widehat{H}_1\,,\, \widehat{H}_2\,,\, \widehat{Q}\,,\,
\widehat{L} \Big)\ \to\ 
\omega\, \Big(\widehat{H}_1\,,\, \widehat{H}_2\,,\, \widehat{Q}\,,\,
\widehat{L} \Big)\;,\qquad
\end{eqnarray}
whereas the remaining iso-singlet chiral superfields, $\widehat{U}$,
$\widehat{D}$ and $\widehat{E}$, do not transform.
Equation~\eqref{eq:Rsyms} implies:
${\cal \widehat{W}}\ \to\ \omega^2\,{\cal \widehat{W}}$.  If
$\omega^2 = 1$, the discrete transformations stated
in~\eqref{eq:Rsyms} give rise to a global ${\bf Z}_2$ symmetry, which
is automatically satisfied by the complete effective superpotential
${\cal W}_{\rm eff}$ in~\eqref{eq:Weff} and the minimal Kaehler
potential ${\cal K}$ in~\eqref{eq:Kmin}. For $\omega^2 \neq 1$,
however, the super\-potential ${\cal W}$ is charged and
\eqref{eq:Rsyms} represents a non-trivial discrete $R$ symmetry, which
is maintained by an appropriate rotation of the Grassmann-valued
coordinates of the SUSY space.

We may now exploit this discrete $R$ symmetry in order to suppress
lower powers of the non-renormalizable operators in the effective
superpotential ${\cal \widehat{W}}$ given in~\eqref{eq:Weff}, as well
as the respective $A_{2n}$ terms induced by ${\cal \widehat{W}}$. Given that
$\widehat{H}_1\widehat{H}_2 \to \omega^2 \widehat{H}_1\widehat{H}_2$
under the discrete $R$-symmetry transformations in~\eqref{eq:Rsyms},
we may now require that
\begin{equation}
\label{eq:omega}
\omega^{2n}\ =\ \omega^2\;,
\end{equation}
for $n > 2$. Note that for $n = 1, 2$, no non-trivial restrictions on
the form of ${\cal \widehat{W}}$ will arise. 

Let us therefore turn our attention to the case with $n=3$
in~\eqref{eq:omega}. This leads to a scenario realizing the discrete
$R$ symmetry ${\bf Z}^R_4$, with $\omega^4 = 1$ and
$\omega^2 = -1\neq 1$. In this case, ${\cal \widehat{W}}$ takes on the form:
\begin{eqnarray}
\label{eq:Wrho6}
\nonumber {\cal \widehat{W}}\ & = &\  {\cal \widehat{W}}_0\: +\: \mu\, \widehat{H}_1
\widehat{H}_2\: 
+\: \frac{\rho_6}{6}\,\frac{(\widehat{H}_1\widehat{H}_2)^3}{M^3}\\
&&+\: \frac{\rho_{10}}{10}\, \frac{(\widehat{H}_1\widehat{H}_2)^5}{M^7}\
+ \ \dots 
\end{eqnarray}
In such a minimal SUGRA framework with $R$ symmetry, the induced
SUSY-breaking potential for the Higgs sector is expected to be of the
form~\cite{Nilles:1983ge}:
\begin{eqnarray}
\label{eq:Vbr6}
\nonumber V^H_{6,{\rm br}}\ &=&\ \bigg( B \mu\, H_1H_2
+ A_6\,\frac{(H_1H_2)^3}{M^3}\\
&&+\: A_{10}\, \frac{(H_1H_2)^5}{M^7}
+  \dots \bigg)\ +\ {\rm H.c.}\quad
\end{eqnarray}
As before, we assume for simplicity that the soft SUSY-breaking mass
parameters~$m^2_{1,2}$ are small, i.e.~$m^2_{1,2} \ll B\mu$, so that
they can be ignored. Likewise, we assume that only the leading
$\rho_6$-coupling and the $A_6$ term are sizeable and so relevant.
In this case, along the $D$-flat direction~\eqref{eq:Dflat}, the
scalar potential for $\phi > M_{\cal S}$ will
acquire the simple form
\begin{eqnarray}
\label{eq:VHiggs6}
\nonumber V_6(\phi)\ &=&\ e^{\phi^2/M^2_{\rm Pl}}\, \bigg[ - \frac{m^2}{2}\,\phi^2 +
\frac{{\rm Re}(e^{3i\xi}A_6)}{4M}\, \frac{\phi^6}{M^2}\\
&&+\ \frac{|\rho_6|^2}{32}\, \frac{\phi^{10}}{M^6}\, 
\bigg( 1 + \frac{9}{72}\,\frac{\phi^2}{M^2_{\rm Pl}} +
\frac{1}{72}\,\frac{\phi^4}{M^4_{\rm Pl}} \bigg)\bigg]\, .\qquad\qquad
\end{eqnarray}
In fact, this last result can be generalized to a discrete
$R$ symmetry ${\bf Z}^R_{2n-2}$, with $\omega^{2(n -1)} =~1$ and
$n \ge 3$. In this case, the leading form of the scalar
potential~$V_{2n}$ for $\phi > M_{\cal S}$ becomes
\begin{eqnarray}
\label{eq:Vn}
V_{2n}(\phi)\ &=&\ e^{\phi^2/M^2_{\rm Pl}}\,\bigg[ - \frac{m^2}{2}\,\phi^2\:
           +\: 
\frac{{\rm Re}(e^{n\,i\xi}A_{2n})}{2^{n-1}\,M}\,
             \frac{\phi^{2n}}{M^{2(n-2)}}\nonumber\\
&&+\ \frac{|\rho_{2n}|^2}{2^{2n-1}}\, \frac{\phi^{2(2n-1)}}{M^{2(2n-3)}} 
\bigg( 1 + \frac{4n-3}{2(2n)^2}\,\frac{\phi^2}{M^2_{\rm
   Pl}}\nonumber\\
&& +\: \frac{1}{2(2n)^2}\,\frac{\phi^4}{M^4_{\rm Pl}} \bigg) \bigg]\, .\qquad
\end{eqnarray}
In the above, we have also neglected all small terms that are proportional
to $|\mu |/M$. If~$A_{2n} > 0$, the proper harmful
$D$-flat direction is obtained for $\xi = \pi/n$, leading to the smallest
negative coefficient for the $\phi^{2n}$ operator in~\eqref{eq:Vn},
since~${{\rm Re}(e^{n\,i\xi}A_{2n}) = - A_{2n} < 0}$.

Before concluding this section, a number of important remarks are in order.
\begin{itemize}
	
	\item[(i)] It is essential that a discrete $R$ symmetry is employed
	and not a continuous one, in order to eliminate the harmful
	Planck-scale operators~$(H_1H_2)^n/M^{2n-3}$ in~\eqref{eq:Vbr}. At
	first instance, a continuous U(1) $R$ symmetry would have forbidden
	all these operators in the scalar potential for $n\ge 2$. However,
	this would have led to a non-convex potential along the $D$-flat
	direction, thereby exacerbating the problem of stability of the
	electroweak vacuum.
	
	\item[(ii)] One may worry that quantum effects may induce further
	non-renormalizable operators${}$ which in turn may alter the leading
	form of the potential in~\eqref{eq:Vn}. However, these operators
	will have higher dimensionality than the leading harmful operator
	and also expected to be typically suppressed by an extra loop factor
	$1/(16\pi^2) \approx 10^{-2}$~\cite{Garbrecht:2006az}. Moreover, in
	Dimensional Regularization (DR), corrections to lower-dimensional
	operators~$\phi^{2n}$ due to seagull graphs resulting from higher
	dimensional operators, such as $\phi^{2n +2}$ and higher, are
	vanishing, if the field direction${}$~$\phi$ is taken to be massless.
	This is indeed a good approximation, given
	that~${m, M_{\cal S} \ll M}$.  Finally, possible corrections to
	SUSY-invariant $F$-term operators in~\eqref{eq:Vn} can only occur
	via two (or even powers) of SUSY-breaking operators.  As a
	consequence, such higher-order corrections will be positive definite
	and so harmless for the stability of the electroweak vacuum.
	
	\item[(iii)] It is known that soft SUSY-breaking bilinear and
	trilinear terms induced by the renormalizable superpotential
	$\widehat{\cal W}_0$ cannot be arbitrary, as they can induce charge
	and colour breaking (CCB) minima~\cite{Dine:1995kz}. Here we simply
	assume that these soft SUSY-breaking operators have been chosen so
	as to avoid the existence of dangerous CCB vacua close to the EW
	scale. As usually done in the formulation of the MSSM, we impose an
	extra non-anomalous matter-parity symmetry, the so-called
	$R$-parity, to ged rid of all renormalizable $(B-L)$-violating
	operators~\cite{Ibanez:1991pr}, and so allow\- for a stable Dark
	Matter candidate and also ensure the longevity of the
	proton${}$. Nevertheless, one may still worry that $(B-L)$-conserving
	non-renormalizable superpotential operators, such as
	$\widehat{Q}\widehat{Q}\widehat{Q}\widehat{L}$ and
	$\widehat{U}\widehat{U}\widehat{D}\widehat{E}$, might produce
	harmful SUSY-breaking operators along their $D$-flat
	direction~\cite{Dine:1995kz,Gherghetta:1995dv}. It is easy to see
	that such operators are not permitted once the discrete $R$ symmetry
	in~\eqref{eq:Rsyms} is imposed, assuming that $\omega^2 \neq 1$.
	However, other harmful non-renormalizable operators of dimension-4
	can survive, such as $\widehat{Q}\widehat{U}\widehat{Q}\widehat{D}$
	and $\widehat{Q}\widehat{U}\widehat{L}\widehat{E}$~\cite{Dine:1995kz,Gherghetta:1995dv}. In
	the present analysis, we have assumed that their SUSY-breaking
	$A$-terms are small enough to trigger transitions into a CCB vacuum,
	along their $D$-flat direction. Alternatively, one may extend the
	MSSM by adding an extra singlet and choosing other sets of discrete
	$R$ symmetries along the lines
	of~\cite{Panagiotakopoulos:2000wp}. Such explorations lie beyond the
	scope of the present study.
	
	\item[(iv)] Likewise, we did not consider explicit
	string-theoretic embeddings of the SUGRA scenarios\- under
	investigation. In this respect, we did not address the issue
	of whether all discrete $R$ symmetries ${\bf Z}^R_{2n-2}$ can
	be made non-anomalous with respect to the SM-gauge and
	gravitational interactions~\cite{Krauss:1988zc,Ibanez:1991pr,
		Banks:1991xj,Kallosh:1995hi}. Instead, we simply note that
	for the class of the effective superpotentials
	${\widehat {\cal W}}$ considered in~\eqref{eq:Weff}, one can
	show~\cite{Byakti:2017igi} that all $R$ symmetries have no
	mixed anomalies with the SU$(3)_c$ gauge group.  Apart
	from~${\bf Z}^R_4$~\cite{Byakti:2017igi}, higher $R$
	symmetries might require the addition of exotic colourless matter\- to
	avoid mixed anomalies with the SU$(2)_L$ and U(1)$_Y$ groups
	of the SM. Finally${}$, gravitational instantons may also break
	anomalously $R$ symmetries. But, there is no
	consensus~\cite{Witten:1985xe} whether these are relevant and
	to what extent they should be included in the path
	integral. Because of the lack of accurate quantitative
	estimates of the effect of gravitational instantons, we will
	assume\- that any instanton-induced violation of $R$
	symmetries, or explicit violations of these through string
	compactifications, are subdominant and can therefore be
	ignored for the purpose of this study.
	
\end{itemize}
In the next section, we will use the leading form of the SUGRA-derived
potential $V_{2n} (\phi )$ in~\eqref{eq:Vn}, for field values
$\phi > M_{\cal S}$, in order to assess the stability of the EW vacuum
against the presence of harmful Planck-scale suppressed operators.

\section{EW Vacuum Stability in SUGRA Models}
\label{sec:sspotential}

In this section, we will analyze the stability of the EW vacuum, upon
minimally embedding\- the SM into an effective SUGRA theory that
happens to predict the leading form of Planckian NP. We will consider
a SUGRA-derived extension of the SM effective potential~$V_{2n}(\phi)$
for Higgs field values~$\phi$ above the soft SUSY-breaking
scale~$M_{\cal S}$. This means that for $\phi > M_{\cal S}$, we will
adopt the leading form of the SUGRA potential $V_{2n} (\phi )$
of~\eqref{eq:Vn}. Instead, for $\phi < M_{\cal S}$, the SM effective
potential~$V_{\rm SM}(\phi)$ given in~\eqref{potential} will be
regarded to be an accurate approximation of the theory,
i.e.~$V_{2n} (\phi < M_{\cal S} ) = V_{\rm SM}(\phi)$.  Given that the
reduced Planck mass~$M_{\rm Pl}$~[cf.~\eqref{MPlanck}] becomes the
relevant mass scale in SUGRA, all mass parameters will be given in
$M_{\rm Pl}$ units. To simplify further our analysis, we identify the
scale~$M$ in~(\ref{eq:Vn}) with $M_{\rm Pl}$, i.e.~$M = M_{\rm Pl}$.

As for the soft SUSY-breaking scale $M_{\cal S}$, we consider two
different scenarios that realize: (i) a very large
$M_{\cal S} = 10^{9}$~TeV; (ii) a relatively low
$M_{\cal S} = 10$~TeV. In all scalar potentials~$V_{2n} (\phi )$, we
select the flat direction for which the CP-odd phase $\xi$
in~\eqref{eq:Dflat} is given by $\xi = \pi/n$.  This gives rise to a
harmful operator $\phi^{2n}/M^{2n-4}_{\rm Pl}$ which has the largest negative
contribution to~$V_{2n} (\phi )$. Furthermore, for the induced
SUSY-breaking trilinears~$A_{2n}$, we assume that they take the following
four discrete values:
\begin{equation}
\label{eq:A2n}
A_{2n}\ =\ M_{\rm Pl}\,,\quad M_{\rm Pl}/5\,,\quad  M_{\rm Pl}/10\,,\quad M_{\rm
	Pl}/50\; .
\end{equation}
Finally, we set all superpotential couplings $\rho_{2n} = 1$, for
simplicity.

\subsection{SUGRA Scenarios with  {\boldmath $M_{\cal S}=10^{9}$}~TeV}

We will first consider a minimal SUGRA scenario with
$M_{\cal S}=10^{9}$~TeV.  The results of our analysis are exhibited in
Table~\ref{tab:MS10to9}, for different values of $n$ corresponding to
the SUGRA potentials $V_{2n}(\phi)$~[cf.~\eqref{eq:Vn}]. In detail,
Table~\ref{tab:MS10to9} shows the value of the AdS vacuum
energy~$V_{\rm min} \equiv V_{2n}(\phi_{\rm min})$ at the Planckian
AdS vacuum~$\phi_{\rm min}$, the field
values, $\phi_0^{\rm \, flat}$ and $\phi_0^{\rm \, curved}$, as
determined at the center of the bounce (with ${\phi_0 \equiv \phi_b(r\!=\!0)}$), as
well as the EW vacuum lifetimes~$\tau^{\rm flat}$ and
$\tau^{\rm curved}$ (in $T_U$ units) for a flat and a curved spacetime
background, respectively.  A~key theoretical parameter in our analysis
is the SUSY-breaking trilinear coupling~$A_{2n}$, which takes four
representative values as stated~in~\eqref{eq:A2n}.

\begin{table*}[t!]
	\renewcommand\arraystretch{1.5}
	\centering
	\begin{tabular}{|c|c|ccccll|}
		\hline
		\hspace{2mm}$n$\hspace{2mm}  & $A_{2n}$ & $V_{\rm min}$ \qquad & $\phi_{\rm min}$ \qquad &
		$\phi_0^{\rm \, flat}$ \qquad & $\phi_0^{\rm \, curved}$ \qquad &
		$\tau^{\rm flat}$
		\qquad &
		$\tau^{\rm curved}$\\
		\hline\hline
		$2$ & 1 & $-4.1791$ \qquad & $1.4310$ \qquad & $1.4281$ \qquad
		&  $1.4253$ \qquad & $10^{-238}$ \qquad & $10^{-238}$ \\ 
		$3$ & 1 & $-5.1768$ \qquad &  $1.4308$ \qquad & $1.4308$ \qquad 
		& $1.4308$ \qquad  & $10^{-238}$ \qquad & $10^{-237}$\\
		$4$  & 1 & $-5.6986$ \qquad & $1.4264$ \qquad & $1.4264$ \qquad 
		& $1.4264$ \qquad & $10^{-238}$ \qquad & $10^{-236}$ \\ 
		\hline\hline
		$2$ & 1/5 & $-0.0133$ \qquad & $0.7161$ \qquad & $0.0021$ \qquad 
		& $0.0019$ \qquad & $10^{-200}$ \qquad & $10^{-200}$ \\  
		$3$ & 1/5 & $-0.0401$ \qquad & $0.9790$ \qquad & $0.9787$\qquad 
		& $0.9786$ \qquad & $10^{-146}$ \qquad & $10^{-135}$ \\ 
		$4$ & 1/5 & $-0.0669$ \qquad & $1.0991$ \qquad & $1.0991$ \qquad 
		& $1.0991$ \qquad & $10^{-129}$ \qquad & $10^{-104}$ \\ 
		\hline\hline
		$2$ & 1/10 & $-0.0014$ \qquad & $0.5122$ \qquad & $0.0013$ \qquad 
		& $0.0013$ \qquad & $10^{-170}$ \qquad & $10^{-170}$ \\
		$3$ & 1/10 & $-0.0057$ \qquad & $0.8268$ \qquad & $0.8262$ \qquad 
		& $0.8261$ \qquad & $10^{75}$ \qquad & $10^{100}$ \\ 
		$4$ & 1/10 & $-0.0108$ \qquad & $0.9809$ \qquad & $0.9809$ \qquad 
		& $0.9809$ \qquad & $10^{193}$ \qquad & $10^{260}$ \\ 
		\hline\hline
		$2$ & 1/50 & $-9.8 \!\times\! 10^{-6}$ \qquad & $0.2307$ \qquad 
		& $0.0008$ \qquad & $0.0008$ \qquad & $10^{61}$ \qquad & $10^{61}$ \\
		$3$ & 1/50 & $-0.00007$ \qquad & $0.5554$ \qquad 
		& $0.5543$ \qquad & $0.5543$ \qquad & $10^{4205}$ \qquad & $10^{4354}$ \\
		$4$ & 1/50 & $-0.00018$ \qquad & $0.7519$ \qquad & $0.7519$ \qquad 
		& $0.7519$ \qquad & $10^{8317}$ \qquad & $10^{9056}$ \\ 
		\hline
	\end{tabular}
	\caption{\em Numerical estimates of the AdS vacuum energy~$V_{\rm min}$ at the AdS vacuum~$\phi_{\rm min}$, the field values~$\phi_0 \equiv \phi_b(0)$ at the
		center of the bounce, the EW vacuum lifetimes~$\tau$ (in $T_U$ units) for a flat and a
		curved spacetime background, in SUGRA scenarios with harmful
		operators~$\phi^{2n}/M^{2n-4}$~[cf.~\eqref{eq:Vn}]. The input parameters for such
		scenarios are:   ${M_{\cal S}=10^{9}}\,$TeV, $M=M_{\rm Pl}$,
		$\rho_{2n} = 1$, while $A_{2n}$ takes  the four discrete
		values given in~\eqref{eq:A2n}. All energy scales are given in units of  the
		reduced Planck mass~$M_{\rm Pl}$.} \label{tab:MS10to9}
\end{table*} 

From Table~\ref{tab:MS10to9}, we observe that for
$A_{2n} = M_{\rm Pl}$, no noticeable stabilizing effect on the EW
vacuum was found, notwithstanding the presence of gravity and 
the induced curved background metric. In fact, we have checked that
$\tau^{\rm flat} \sim \tau^{\rm curved}$, for very high values of $n$
as well.  Although this result may seem counter-intuitive, it certainly
implies that the protective mechanism presented in
Section~\ref{sec:SUGRA} appears to be ineffective to assure the
stability of our EW vacuum in this case.

As $A_{2n}$ assumes smaller values as shown in
Table~\ref{tab:MS10to9}, e.g.~$A_{2n}= M_{\rm Pl}/5$, we notice that
unlike $n=2$, the lifetime of the EW vacuum, $\tau^{\rm flat}$ and
$\tau^{\rm curved}$ evaluated separately for a flat and a curved
spacetime metric, gets prolonged, as expected. For all the scenarios
with $n=2$, the destabilizing effect of the negative $\phi^4$
potential term is so strong that even the inclusion of gravity can no
longer alter the value of~$\tau$. Otherwise, we anticipate on general
grounds that the inclusion of gravity will increase the stability of
the EW vacuum for all scenarios~$n\ge 3$. However, for the scenario
with~$A_{2n}= M_{\rm Pl}/5$, all low order harmful operators with
$n = 2,\, 3,\, 4$ lead to lifetimes $\tau \ll T_U$, as can be seen from
Table~\ref{tab:MS10to9}. When $A_{2n}$ becomes even smaller,
i.e.~$A_{2n} = M_{\rm Pl}/10$ and $A_{2n} = M_{\rm Pl}/50$, a quicker
stabilization of the EW vacuum is achieved and the predicted
tunnelling time~$\tau$ becomes much larger than the age of the
Universe~$T_U$, for all scenarios with $n\ge 3$ and $n \ge 2$,
respectively. This result is in agreement with the discussion
presented in Section~\ref{sec:PlanckNP}, since the negative
$A_{2n}$-dependent contribution of the harmful operators to the
potentials~$V_{2n}$ becomes less significant for scenarios with lower
values~$A_{2n}$.

\begin{figure}[t!] 
	\centering
	${}$\hspace{-2mm}\includegraphics[width=0.49\textwidth]{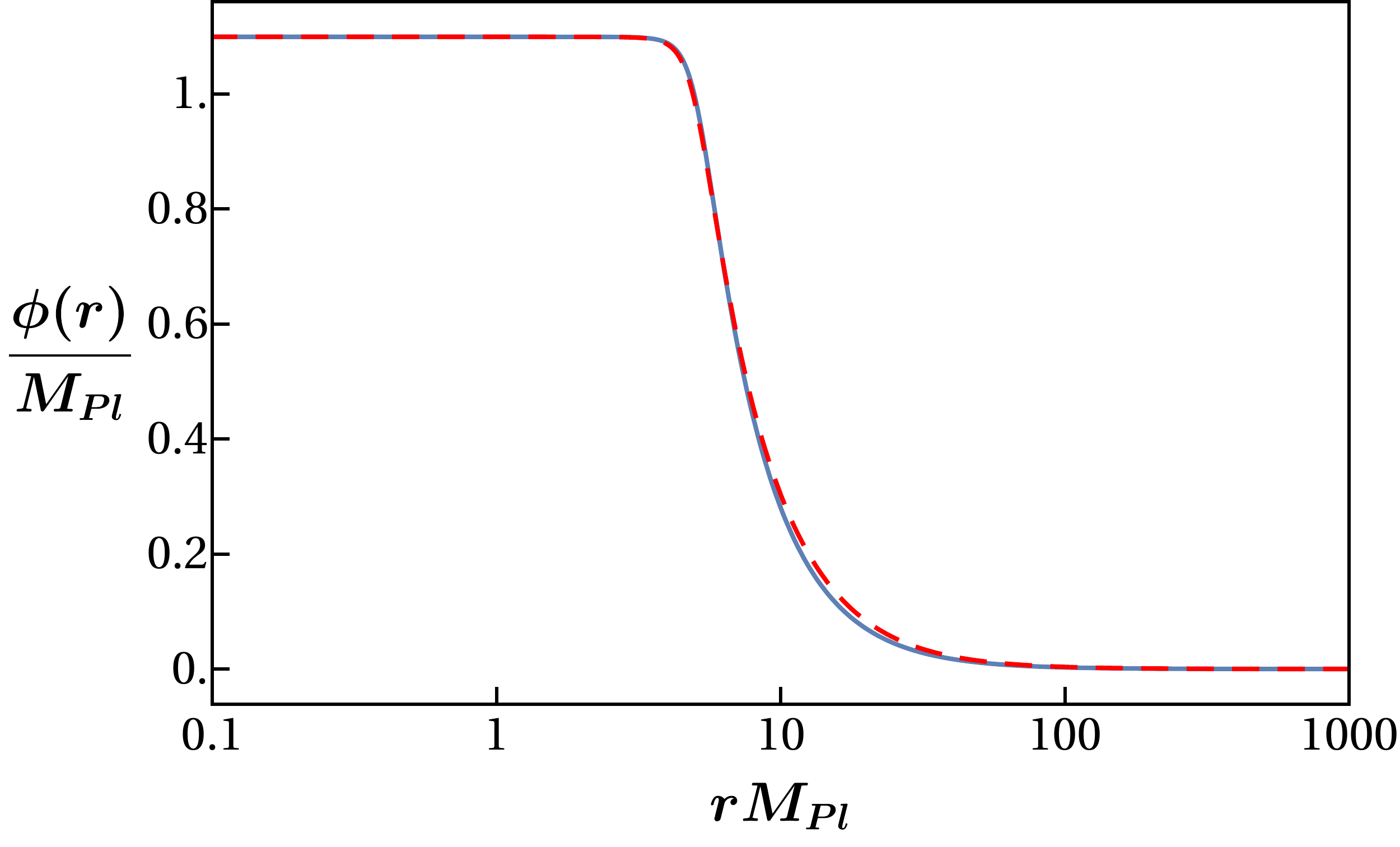}\\[7mm]
	\includegraphics[width=0.46\textwidth]{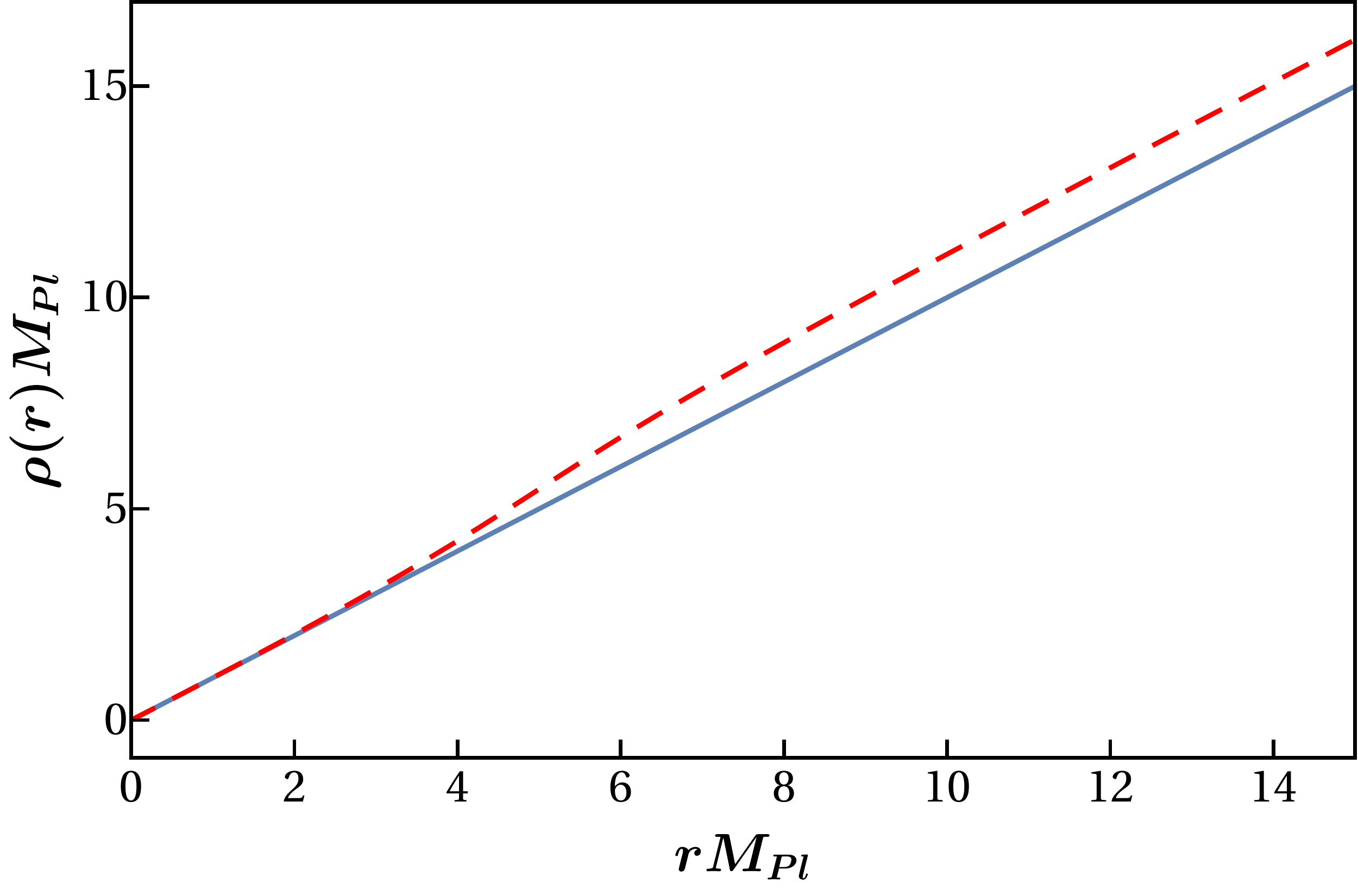}
	\caption{\em The bounces $\phi(r)$ (upper panel) and the curvatures
		$\rho(r)$ (lower panel) for the potential (\ref{eq:Vn}) with
		$A_{2n}= M_{\rm Pl}/5$ and $n=4$, for a flat (solid blue)
		and a curved (dashed red) spacetime.}\label{fig:fignew}
\end{figure}

Finally, it is interesting to observe that as $n$ increases, the AdS
vacuum $\phi_{\rm min}$ and the bounces, $\phi_{0}^{\rm flat}$ and
$\phi_{0}^{\rm curved}$ at~$r=0$, all start to converge towards the
same value:
${\phi_{\rm min} = \phi_{0}^{\rm flat} =\phi_{0}^{\rm curved} }$.  In
the same context, we have verified that the whole radial
profile${}$~$\phi^{\rm flat}(r)$ will start to coincide with that
of~$\phi^{\rm curved}(r)$. In fact, the difference between
$\tau^{\rm flat}$ and $\tau^{\rm curved}$ found
in~Table~\ref{tab:MS10to9} will result from the two actions of the
bounce solutions [cf.~\eqref{eq:Sflat} and \eqref{gravaction}],
\begin{equation*}
S_b^{\rm flat}\ =\ -2\pi^2\int_0^\infty dr\, r^3\, V(\phi^{\rm flat})
\end{equation*}
and
\begin{equation*}
S_b^{\rm curved}\ =\ -2\pi^2\int_0^\infty dr\, \rho^3\, V(\phi^{\rm
	curved})\;,
\end{equation*}
which determine the tunnelling exponent~$B$ in~\eqref{tau}. Hence, the
deviation of the curvature of the metric, $\rho = \rho(r)$ (curved
spacetime), from the respective flat one, $\rho^{\rm flat}(r) = r$,
will control the difference in the predictions for $\tau^{\rm flat}$
versus $\tau^{\rm curved}$. 

In~Figure${}$~\ref{fig:fignew}, we give a concrete example, where we
plot the bounces~$\phi(r)$ (upper panel) and the curvatures~$\rho(r)$
(lower panel) for the potential~(\ref{eq:Vn}) with $n =4$ and
$A_{2n}= M_{\rm Pl}/5$, for a flat (solid blue line) and a curved
(dashed red line) background metric. We see that while the two bounce
solutions for $\phi(r)$ practically coincide, the corresponding ones
for $\rho(r)$ differ from one another, thereby causing the prediction
for~$\tau^{\rm flat}$ to significantly deviate from that
for~$\tau^{\rm curved}$.

\begin{table*}[t!]
	\renewcommand\arraystretch{1.5}
	\centering
	\begin{tabular}{|c|c|ccccll|}
		\hline
		\hspace{2mm}$n$\hspace{2mm}  & $A_{2n}$ & $V_{\rm min}$ \qquad & $\phi_{\rm min}$ \qquad &
		$\phi_0^{\rm \, flat}$ \qquad & $\phi_0^{\rm \, curved}$ \qquad &
		$\tau^{\rm flat}$
		\qquad &
		$\tau^{\rm curved}$\\
		\hline\hline		
		$2$ & 1 & $-4.1791$ \qquad & $1.4310$ \qquad & $1.4281$ \qquad 
		& $1.4253$ \qquad & $10^{-238}$ \qquad & $10^{-238}$ \\ 		
		$3$ & 1 & $-5.1768$ \qquad & $1.4308$ \qquad & $1.4308$ \qquad 
		& $1.4308$ \qquad & $10^{-238}$ \qquad & $10^{-237}$ \\ 		
		$4$ & 1 & $-5.6986$ \qquad &  $1.4264$ \qquad & $1.4264$ \qquad 
		& $1.4264$ \qquad & $10^{-238}$ \qquad & $10^{-236}$ \\ 
		\hline\hline
		$2$ & 1/5 & $-0.0133$ \qquad & $0.7161$ \qquad & $2.24\!\times\! 10^{-7}$ \qquad 
		& $2.04\!\times\! 10^{-7}$ \qquad & $10^{-184}$ \qquad & $10^{-184}$ \\ 	
		$3$ & 1/5  & $-0.0401$ \qquad  & $0.9790$ \qquad & $0.9787$ \qquad 
		& $0.9786$ \qquad & $10^{-146}$ \qquad & $10^{-135}$ \\ 		
		$4$ & 1/5  & $-0.0669$ \qquad & $1.0991$ \qquad & $1.0991$ \qquad 
		& $1.0991$ \qquad & $10^{-129}$ \qquad & $10^{-104}$ \\ 
		\hline\hline
		$2$ & 1/10 & $-0.0014$ \qquad & $0.5123$ \qquad & $1.49\!\times\!10^{-7}$ \qquad 
		& $1.47\!\times\!10^{-7}$ \qquad &
		$10^{-154}$ \qquad & $10^{-154}$ \\ 
		$3$ & 1/10 & $-0.0057$ \qquad & $0.8268$ \qquad & $0.8262$ \qquad 
		& $0.8261$ \qquad & $10^{76}$ \qquad & $10^{100}$ \\ 		
		$4$ & 1/10 & $-0.0108$ \qquad & $0.9809$ \qquad &  $0.9809$ \qquad 
		& $0.9809$ \qquad & $10^{218}$ \qquad & $10^{260}$ \\ 
		\hline\hline
		$2$ & 1/50 & $-9.8\!\times\! 10^{-6}$ \qquad & $0.2307$ \qquad & 
		$1.10\!\times\! 10^{-7}$ \qquad 
		& $1.10\!\times\!10^{-7}$ \qquad & $10^{76}$ \qquad & $10^{76}$ \\ 
		$3$ & 1/50 & $-0.00008$ \qquad & $0.5554$ \qquad & $0.5543$ \qquad 
		& $0.5543$ \qquad & $10^{4196}$ \qquad & $10^{4354}$ \\ 
		$4$ & 1/50 & $-0.00018$ \qquad & $0.7519$ \qquad & $0.7519$ \qquad 
		& $0.7519$ \qquad & $10^{8006}$ \qquad & $10^{9056}$ \\ 
		\hline
	\end{tabular}
	\caption{\em The same as in Table~\ref{tab:MS10to9}, but with $M_{\cal S}=10\,$TeV.}\label{tab:MS10}
\end{table*} 

\subsection{SUGRA Scenarios with  {\boldmath $M_{\cal S}=10$}~TeV}

We will now study a class of minimal SUGRA scenarios with a soft
SUSY-breking scale~$M_{\cal S}=10$~TeV. Such scenarios are better
motivated, in the sense that they require a much smaller degree of
fine tuning for solving the infamous gauge hierarchy
problem. Otherwise, all other theoretical parameters take the same
values as before. This exercise will allow us to probe the sensitivity
of our results to~$M_{\cal S}$. Table~\ref{tab:MS10} summarizes the
findings of our analysis.

As was the case for the SUGRA scenarios with $M_{\cal S}=10^{9}$~TeV,
we find similar features for those with~$M_{\cal S}=10$~TeV. As
before, we obtain that for any fixed value of~$A_{2n}$, the EW vacuum
lifetime~$\tau$ will increase with $n$. Likewise, $\tau$ will also
increase, as~$A_{2n}$ decreases. Similarly, we notice that for $n=2$,
the destabilizing effect of the negative $\phi^4$ potential term is
strong enough to counter-act the respective stabilising effect thanks
to gravity. As a consequence, the predictions for~$\tau$ turn out not
to depend on the choice of the background metric. As before, we
observe that for ${A_{2n}= M_{\rm P}/50}$, the EW vacuum lifetime
gets adequately prolonged, becoming much larger than $T_U$, already
from $n=2$ and on.

If we compare the results presented in Table~\ref{tab:MS10} to those
in Table~\ref{tab:MS10to9}, we will observe that the tunnelling
times~$\tau^{\rm flat}$ and~$\tau^{\rm curved}$ are rather comparable,
especially when ${n\ge 3}$.  Evidently, this suggests that even if
gravity is taken into account, the bounce solutions to~\eqref{meq}
seem to be insensitive to the matching of the SM effective potential
$V_{\rm SM}(\phi)$ to the SUGRA potential~$V_{2n}(\phi)$ for a wide
range of $\phi$ values: $\phi=10-10^{9}$~TeV.

In order to gain further insight into this point, we consider a
variant effective scalar potential~$\widetilde{V}_{2n}(\phi )$. To be
precise, for $\phi < 10$~TeV, we set
$\widetilde{V}_{2n}(\phi ) = V_{\rm SM}(\phi)$, and for
$\phi > 10^{9}$~TeV, $\widetilde{V}_{2n}(\phi ) =
V_{2n}(\phi)$.
However, between the field values $\phi = 10$~TeV and
$\phi = 10^{9}$~TeV, the new potential~$\widetilde{V}_{2n}(\phi )$ is
assumed to follow an interpolating straight line. Interestingly
enough, the tunnelling times~$\tau$, as well as the other parameters
shown in Table~\ref{tab:MS10}, come out to be close to the
corresponding ones obtained when $V_{\rm SM}(\phi)$ or $V_{2n}(\phi)$
are used as an interpolation from 10 to $10^{9}$~TeV. Consequently,
the bounce solutions turn out to be insensitive to the shape of the
potential in the above range of~$\phi$. This observation explains the
robustness and the independence of our results for a wide range of
soft SUSY-breaking scales: $M_{\cal S} =10-10^{9}$~TeV.

\section{Conclusions}
\label{sec:Concl}

We have studied the stability of the EW vacuum along the radial
direction~$\phi$ of the Higgs doublet in the presence of Planck-scale
suppressed operators of the form~$\phi^{2n}/M^{2n-4}$, where $M$ is of
order the Planck mass $M_{\rm P}$. Such operators can no longer be
excluded in the presence of gravity, as they could in principle be
generated by quantum gravity effects. If these operators contribute
with a negative sign to the SM effective potential, they will in
general have a destabilizing effect on the EW vacuum, and therefore we
have called them~{\em harmful} throughout this study.

After briefly describing the theoretical framework for evaluating the
lifetime~$\tau$ of the EW vacuum in the SM, we have employed it to
re-establish the leading-order results for~$\tau$ known in the
literature (in $T_U$ units), i.e.~${\tau_{\rm SM} \sim 10^{639}}$ and
${\tau_{\rm SM} \sim 10^{661}}$ for a Minkowski and
$\mathbb{O}(4)$-symmetric gravitational background metric,
respectively. In~this theoretical framework, we have then evaluated
the lifetime~$\tau$ of the EW vacuum\- for simple scenarios of
Planckian NP with convex potentials in the presence of harmful
operators. For $n=3$, $M = M_{\rm P}$ and $c_2 = c_1 = -2$
[cf.~\eqref{new}], we recover the recent results quoted
in~\cite{Bentivegna:2017qry}, thereby corroborating the importance of the harmful
Planck-scale operators on~$\tau$, even in the presence of a curved
spacetime metric, i.e.~$\tau/T_U \sim 10^{-122}$. For such scenarios,
we have found that longevity of the EW vacuum requires $n \ge 4$,
leading to $\tau \stackrel{>}{{}_\sim} \tau_{\rm SM}$. Most
remarkably${}$, for theories with relatively lower scale $M$ of
quantum gravity, e.g.~$M = M_{\rm P}/10$, a safely stable EW vacuum
implies that all harmful operators up to $n= 6$ need to be {\em
	either} accidentally suppressed {\em or} naturally eliminated by the
action of some symmetry.

Besides resorting to {\it ad hoc} accidental suppressions, in this
paper we have explored the possibility${}$ whether the harmful
Planck-scale operators of the form~$\phi^{2n}/M^{2n-4}$ could be
eliminated {\em naturally} to leading order because of the action of
some symmetry. In this context, we have shown how minimal embeddings
of the SM in SUGRA can stabilize the EW vacuum against these harmful
operators up to very high values of the induced SUSY-violating
$A$-couplings $A_{2n}$ and the soft SUSY-breaking
scale~$M_{\cal S}$. The scale~$M_{\cal S}$ may even lie above the
so-called SM meta\-stability scale of~$10^{11}$~GeV.  In particular${}$,
we have explicitly demonstrated how discrete\- $R$ symmetries, such as
${\bf Z}^R_{2n -2}$, could be invoked to suppress the harmful
operators to arbitrary higher powers of~$n$.  In this minimal SUGRA
framework, we~have analyzed different scenarios of Planck-scale
gravitational physics and derived lower limits on the power~$n$ that
will be needed in order to render the EW vacuum sufficiently
long-lived. We have presented numerical estimates for a few
representative scenarios realising a low and high soft SUSY-breaking
scale~$M_{\cal S}$, i.e.~for $M_{\cal S} = 10$~TeV and
$M_{\cal S} = 10^9$~TeV. In all our numerical estimates, the effect of
gravity on the tunnelling time~$\tau$ from the false EW vacuum to the
true Planckian vacuum was carefully considered.

The present study has revealed the severity of a problem for theories
with low-scale quantum gravity. In particular, we have illustrated
that such theories face serious difficulties${}$ in ensuring adequate
longevity of our EW vacuum. These theories may have a string-theoretic
origin~\cite{Horava:1995qa} giving rise to realizations with a lower
effective Planck mass, including models with large compact
dimensions~\cite{Antoniadis:1998ig,Randall:1999ee}. It would
be interesting to analyse the restrictions that can be derived from
the evaluation of~$\tau$ on the fundamental parameters of such
theories.

\acknowledgments

The authors wish to dedicate the present work to the memory of Maria
Krawczyk, acknowledging her warm hospitality at Warsaw U., as well as
her strong interest in this study during the early stages of the project.  We also
thank Elias Kiritsis for useful discussions concerning $R$ symmetries
in string theory.  The~work of VB and FC is carried out within the
INFN project QFT-HEP and is supported in part by the Polish National
Science Centre HARMONIA grant under contract UMO-2015/18/M/ST2/00518
(2016-2019). The work of AP is supported in part by the
Lancaster--Manchester--Sheffield Consortium${}$ for Fundamental
Physics, under STFC research grant ST/L000520/1.

\vfill\eject

\bibliography{Protect_EW_Vacuum_PRD}

\end{document}